\documentclass[prd,aps,amssymb,nofootinbib,showpacs,preprint]{revtex4}

\usepackage{amsmath,amssymb,amsfonts}
\usepackage{graphics,graphicx}
\usepackage[unicode,colorlinks,linkcolor=blue]{hyperref}

\begin{document}
\title{Classical evolution of subspaces}
\author{Yana~Lyakhova}
\affiliation{National Research Nuclear University MEPhI (Moscow Engineering Physics Institute),\\ 115409, Kashirskoe shosse 31, Moscow, Russia}
\author{Arkady~A.~Popov}
\affiliation{N.~I.~Lobachevsky Institute of Mathematics and Mechanics, Kazan  Federal  University, \\ 420008,   
Kremlevskaya  street  18,  Kazan,  Russia}
\author{Sergey~G.~Rubin}
\affiliation{National Research Nuclear University MEPhI (Moscow Engineering Physics Institute),\\ 115409, Kashirskoe shosse 31, Moscow, Russia}
\affiliation{N.~I.~Lobachevsky Institute of Mathematics and Mechanics, Kazan  Federal  University, \\ 420008,   
Kremlevskaya  street  18,  Kazan,  Russia}

\begin{abstract}We study evolution of manifolds after their creation at high energies. Several kinds of gravitational Lagrangians with higher derivatives are considered. 
It is shown analytically and confirmed numerically that an asymptotic growth of the maximally symmetric manifolds depends strongly on their dimensionality. A number of final metrics describing our Universe is quite poor if we limit ourselves with a maximally symmetric extra space. 
We show that the initial conditions can be a reason of nontrivial solutions (funnels) and study their properties.
\end{abstract}

\maketitle

	\section{Introduction}
	
The compact extra spaces is widely used idea. Their inclusion into physical theories helps to move forward on such issues as the grand unification \cite{1998PhLB..429..263A,1999NuPhB.537...47D}, neutrino mass \cite{2002PhRvD..65b4032A}, the cosmological constant problem \cite{2001PhRvD..63j3511S,2003PhRvD..68d4010G} and so on.
Any multi-dimensional model has to lead to the effective 4-dim theory at low energies.  This would imply  relations between the observable four-dimensional physics and a metric of the higher dimensions. 

One of the question remaining not clarified yet is: why specific number of dimensions are compactified and stable while others expand \cite{2007JHEP...11..096G,2002PhRvD..66b4036C,2002PhRvD..66d5029N}?
Which specific property of subspace leads to its quick growth? There are many attempts to clarify the problem, mostly related to introduction of fields other than gravity. It may be a scalar field \cite{2017PhRvD..95j3507K,2007JHEP...11..096G} (most used case), gauge fields \cite{2009PhRvD..80f6004K}. A static solutions can be obtained using the Casimir effect \cite{1984NuPhB.237..397C} or form fields \cite{1980PhLB...97..233F,osti_5368319}. Sometimes one of the subspace is assumed to be FRW space by definition \cite{2013bhce.conf.....B}. Another possibility was discussed in \cite{1984PhRvD..30..344Y,2013GReGr..45.2509B}: it was shown that if the scale factor $a(t)$ of our 3D space is much larger than the growing scale factor $b(t)$ of the extra dimensions, a contradiction with observations can be avoided.

The origin of our Universe is usually related to its quantum creation from the space-time foam at high energies \cite{1994CQGra..11.2483V,2009arXiv0909.2566H}.   The probability of its creation is widely discussed, see e.g. \cite{2013arXiv1311.0220C}. Here we are interested in the subsequent classical evolution of the metrics rather than a calculation of this probability.
Manifolds are nucleated having specific metrics. The set of such metrics is assumed to be very rich. After nucleation, these manifolds evolve classically forming a set of asymptotic manifolds, one of which could be our Universe. In this paper the asymptotic set of the maximally symmetric manifolds with positive curvature is studied in the framework of pure gravity with higher derivatives. We consider models of the $f(R)$ gravity and a more general model acting in 5 and 6 dimensions. No other fields are attracted to stabilize an extra space. 
We have found out that a number of asymptotic solutions is quite limited. This conclusion was confirmed both analytically and numerically. There is a set of initial conditions that lead to a common asymptote of classical solutions. In Section III we have elaborated a method for prediction the asymptotic behavior of metric judging on the specific form of the initial metric.
We also study the funnel solution \cite{2016PhLB..759..622R} as the result of an inhomogeneity of initial metric. 

The gravity with higher derivatives is widely used in modern research despite the internal problems inherent in this approach \cite{2015arXiv150602210W}. Attempts to avoid the Ostrogradsky instabilities are made \cite{2017PhRvD..96d4035P} and extensions of the Einstein-Hilbert action attract much attention.
Promising branch of such models is based on the Gauss-Bonnet Lagrangian and its generalization to the Lovelock gravity. These models were adjusted to obtain differential equations of the second order so that the Ostrogradsky theorem is not dangerous for such models. 

A lot of papers devoted to the $f(R)$-gravity - the simplest extension of the Einstein-Hilbert gravity. Reviews \cite{2010LRR....13....3D}, \cite{2011PhR...509..167C} contain description of the $f(R)$-theories including extension to the Gauss-Bonnet gravity. Examples of research with specific form of the function $f(R)$ can be found in
 \cite{2006GrCo...12..253S}, \cite{2006A&AT...25..447S}. Most of the research assume positive curvature of extra space metric, but as was shown in \cite{2003PhRvL..91f1302T}, hyperbolic manifolds can also be attracted to explain the observable acceleration of the Universe.

In the framework of the gravity with higher derivatives, a variety of regimes with expanding three and contracting extra dimensions has been found in \cite{2017EPJC...77..503P}, \cite{1984PhLB..137..155S}, \cite{1984PhRvD..30.2495S}, \cite{2018EPJC...78..373P},\cite{2007CQGra..24.3713S}. 
The power-law and the exponential analytical behavior of scale-factors are studied in  \cite{2010IJGMM..07..797I}, \cite{2016GrCo...22..329I}. 
Stability of specific extra space metrics is discussed in \cite{2016GrCo...22..329I,2011MPLA...26..805L, 2015PhRvD..92j4017P,2003PhRvD..68d4010G}. The conclusion is that stable metrics do exist but their fraction is quite small.  Here we consider a wider class of metrics depending on the initial conditions. It was found out that different initial metrics can lead to one and the same asymptotic solution.

Throughout this paper we use the conventions for the curvature tensor $R_{ABC}^D=\partial_C\Gamma_{AB}^D-\partial_B\Gamma_{AC}^D+\Gamma_{EC}^D\Gamma_{BA}^E-\Gamma_{EB}^D\Gamma_{AC}^E$
and for the Ricci tensor $R_{MN}=R^F_{MFN}$.
		
	\section{Destiny of subspaces. Exact results. }

    \subsection{Setup and classical equations}

In this section we analyze the classical behavior of the extra space metrics. This can be done on the basis of two well known frames - the Jordan frame and the Einstein one, which are connected by the conformal transformation \cite{2003gr.qc....10112B}. There are intensive debates on the selection of the frame that should be used for appropriate description of the Nature \cite{2018GrCo...24..154B}. Our analysis is mainly based on the Jordan frame.

Let a $D=1+d_1+d_2$-dimensional space-time $T\times M_{d_1}\times M_{d_2}$ has been nucleated due to some quantum processes at high energies. The probability of this process is a subtle point and we do not discuss it in this paper. The entropy growth leads to evolution of subspaces, to those which are the maximally symmetric \cite{Kirillov:2012gy}. In this paper we study the classical evolution of subspaces $M_{d_1}$ and $M_{d_2}$ whose metric
	\begin{equation}\label{metric1}
	ds^2 = dt^2 - e^{2\beta_1(t)}d\Omega_1^2 - e^{2 \beta_2(t)}d\Omega_2^2
	\end{equation}
%
is assumed to be maximally symmetric with a positive curvature. It is supposed that manifolds are born with accidental shape. Subsequently they acquire symmetries due to the entropy growth \cite{Kirillov:2012gy}. We start our study after the process of symmetrization is finished.
In this section, we consider the following action
	\begin{equation}\label{action2}
	S = \frac{m_{D}^{4}}{2}\int d^D Z \sqrt{|g|} f(R),
	\end{equation}
where $R$ is the scalar curvature of a $D$-dimensional space-time. This action appears to be an appropriate tool to study the behavior of the system just after its nucleation.
	
Einstein's equations of this theory are
	\begin{equation}\label{AB}
	-\frac{1}{2}f(R)\delta_A^B + (R_A^B +\nabla_{A}\nabla^{B} - \delta_A^B\square) f_R = 0.
	\end{equation}
Using the results given in Appendix A, we can write the nontrivial equations of this system as
	\begin{eqnarray}\label{11a}
	&& -\frac12 f(R)+f_R [e^{-2\beta_1(t)}(d_1-1)+\ddot{\beta}_1+\dot{\beta}_1(d_1\dot{\beta}_1 +d_2\dot{\beta}_2)] \nonumber \\
    &&+[(1-d_1 )\dot{\beta}_1 - d_2 \dot{\beta}_2  ]f_{RR}\dot{R}-f_{RRR}\dot{R}^2 - f_{RR}\ddot{R} =0,
	\end{eqnarray}
	\begin{eqnarray}\label{22a}
	&& -\frac12 f(R)+f_R[e^{-2\beta_2(t)}(d_2-1)+\ddot{\beta}_2+\dot{\beta}_2(d_1\dot{\beta}_1 +d_2\dot{\beta}_2) ] \nonumber \\
    &&+[(1-d_2 )\dot{\beta}_2 - d_1 \dot{\beta}_1  ]f_{RR}\dot{R}-f_{RRR}\dot{R}^2 - f_{RR}\ddot{R} =0,
	\end{eqnarray}
	\begin{equation}\label{00a}
	-\frac12 f(R)+\left[d_1 \ddot{\beta}_1 +d_2 \ddot{\beta}_2 +d_1{\dot{\beta}_1}^2 + d_2 {\dot{\beta}_2}^2  \right]f_R
	-\left( d_1 \dot{\beta}_1 +d_2 \dot{\beta}_2 \right)f_{RR}\dot{R} =0
	\end{equation}
    in terms of metric \eqref{metric1}.
  Here we have kept in mind $\partial_t f_R=f_{RR}\dot{R}$ and $\partial^2_t f_R=f_{RRR}\dot{R}^2 + f_{RR}\ddot{R}$.
    According to \eqref{R}, the Ricci scalar is
	\begin{eqnarray} \label{Ra}
	&&R=d_1\dot{\beta}_1^2 +  d_2\dot{\beta}_2^2+  d_1\ddot{\beta}_1 + d_2\ddot{\beta}_2
	 + d_1 \left[e^{-2\beta_1(t)}(d_1 -1)+\ddot{\beta}_1
    \right. \nonumber \\ && \left.
    +\dot{\beta}_1 (d_1\dot{\beta}_1 + d_2\dot{\beta}_2)\right] + d_2[e^{-2\beta_2(t)}(d_2 -1)+\ddot{\beta}_2+\dot{\beta}_2(d_1\dot{\beta}_1+d_2\dot{\beta}_2)].
	\end{eqnarray}
For calculations it is convenient to consider the Ricci scalar $R(t)$ as additional unknown function and the definition
\eqref{Ra} as the fourth equation. Three equations of this system (for example, \eqref{11a}, \eqref{22a}, \eqref{Ra})
can be solved with respect to the higher derivatives $\ddot{\beta}_1, \ddot{\beta}_2, \ddot{R}$.
Then substituting $\ddot{\beta}_1$ and $\ddot{\beta}_2$ into equation (\ref{00a}), we obtain the equation
\begin{eqnarray} \label{eq00a}
&&-5 f(R)+\left[5 R -5 d_1(d_1 -1){\dot{\beta}_1}^2  -10 d_1 d_2 \dot{\beta}_1 \dot{\beta}_2
-5 d_2(d_2 -1){\dot{\beta}_2}^2
\right. \nonumber \\ && \left.
 -5 d_1(d_1 -1) e^{-2\beta_1} -5 d_2(d_2 -1) e^{-2\beta_2}\right]f_R
-10 \left[ d_1 \dot{\beta}_1 + d_2 \dot{\beta}_2 \right]f_{RR}\dot{R} =0,
\end{eqnarray}
which plays the role of restricting the solutions  of the coupled second order differential equations.
This can be checked, for example, by writing the set of four equations \eqref{11a}, \eqref{22a}, \eqref{Ra}, \eqref{eq00a}
as an equivalent set of (six) coupled first-order equations plus one algebraic equation.
The equation \eqref{eq00a} reduces to the algebraic transcendental equation, i.e., it is a constraint.
The complete set of initial conditions therefore requires specifying six pieces of information,
namely  $\beta_1 (t_0), \beta_2 (t_0), R(t_0), \dot{\beta}_1 (t_0), \dot{\beta}_2 (t_0)$, and $\dot{R}(t_0)$. These initial conditions are not independent due to equation \eqref{eq00a}. The latter will be used to derive an exact relation between these initial data.

\subsection{ Analysis of numerical solutions and their asymptotes. }

The system of differential equations is highly nonlinear so that one could expect a rich set of its solutions. In this section a set of solutions is  discussed. It will be also shown that solutions and their asymptotes depend on the dimensionality of the extra spaces. To perform numerical simulation one needs to specify the form of the function $f(R)$ and the initial values of the functions $\beta_1(t), \beta_2(t), \dot \beta_1(t), \dot \beta_2(t), \dot R(t)$ at $t=0$. The initial value $R(0)$ is found from auxiliary condition \eqref{eq00a}. Let
\begin{equation}\label{f}
f(R)=a_3R^3 + a_2R^2 + a_1R+a_0
\end{equation}
with the parameter values
\begin{equation}
a_3=1.3, a_2=-2, a_1=1,a_0=0.6.
\end{equation}
Our choice being quite arbitrary doesn't contain small parameters.
Now we are ready to fix the subspace dimensions and perform numerical calculations.
\subsubsection{The case $d_1=3,d_2=3$}

\begin{figure}[ht!]
\includegraphics[width=5cm]{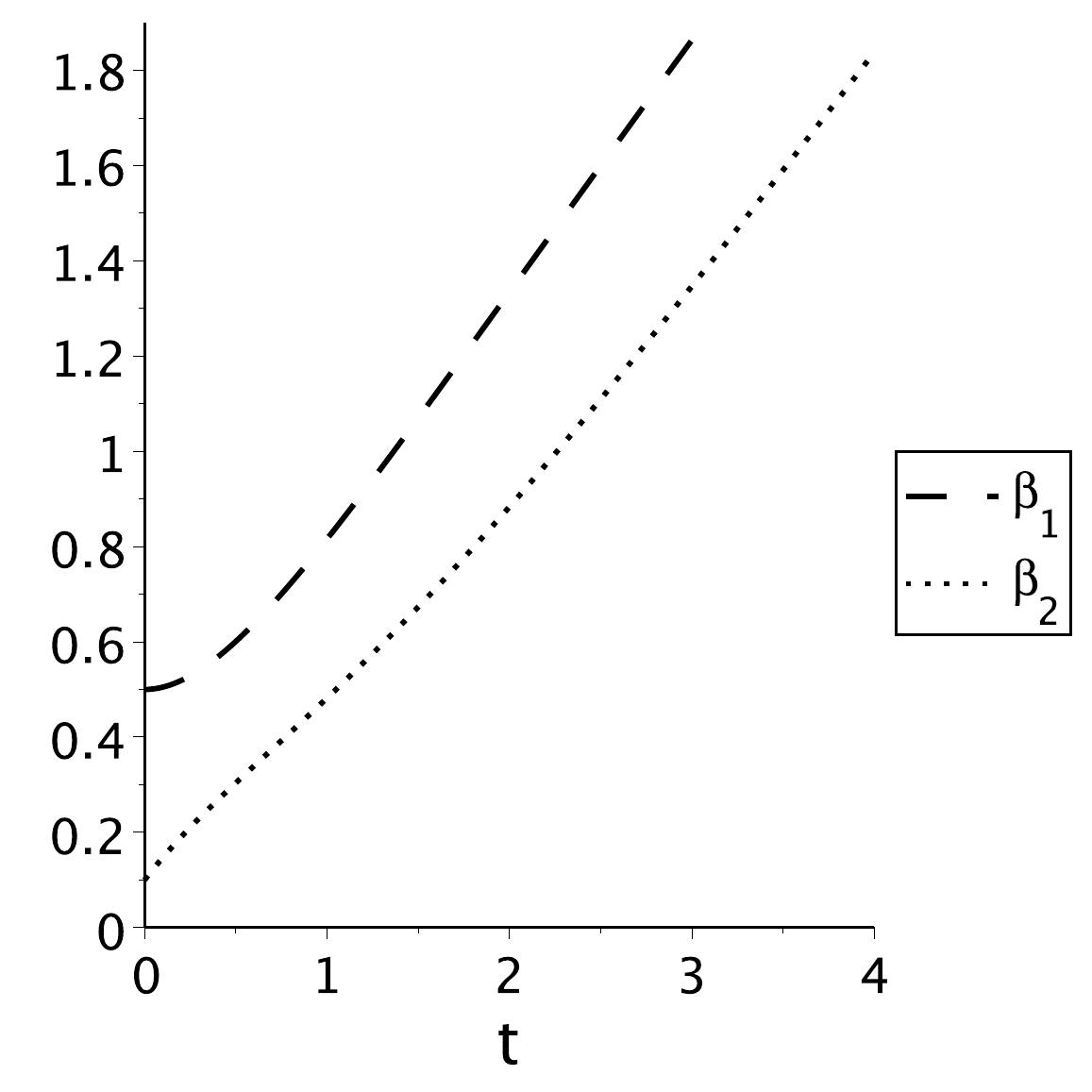}  \quad \includegraphics[width=5cm]{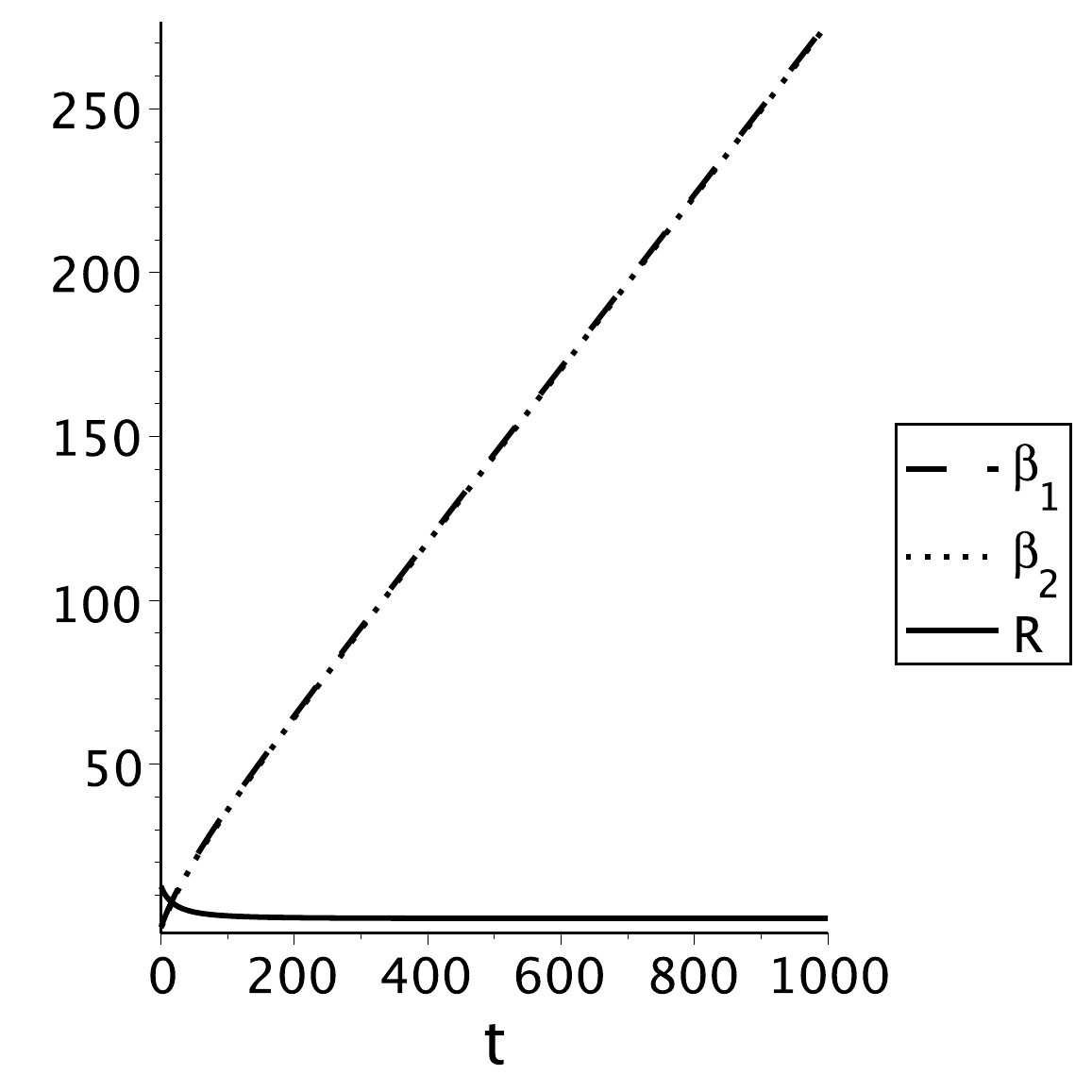}   \quad \includegraphics[width=5cm]{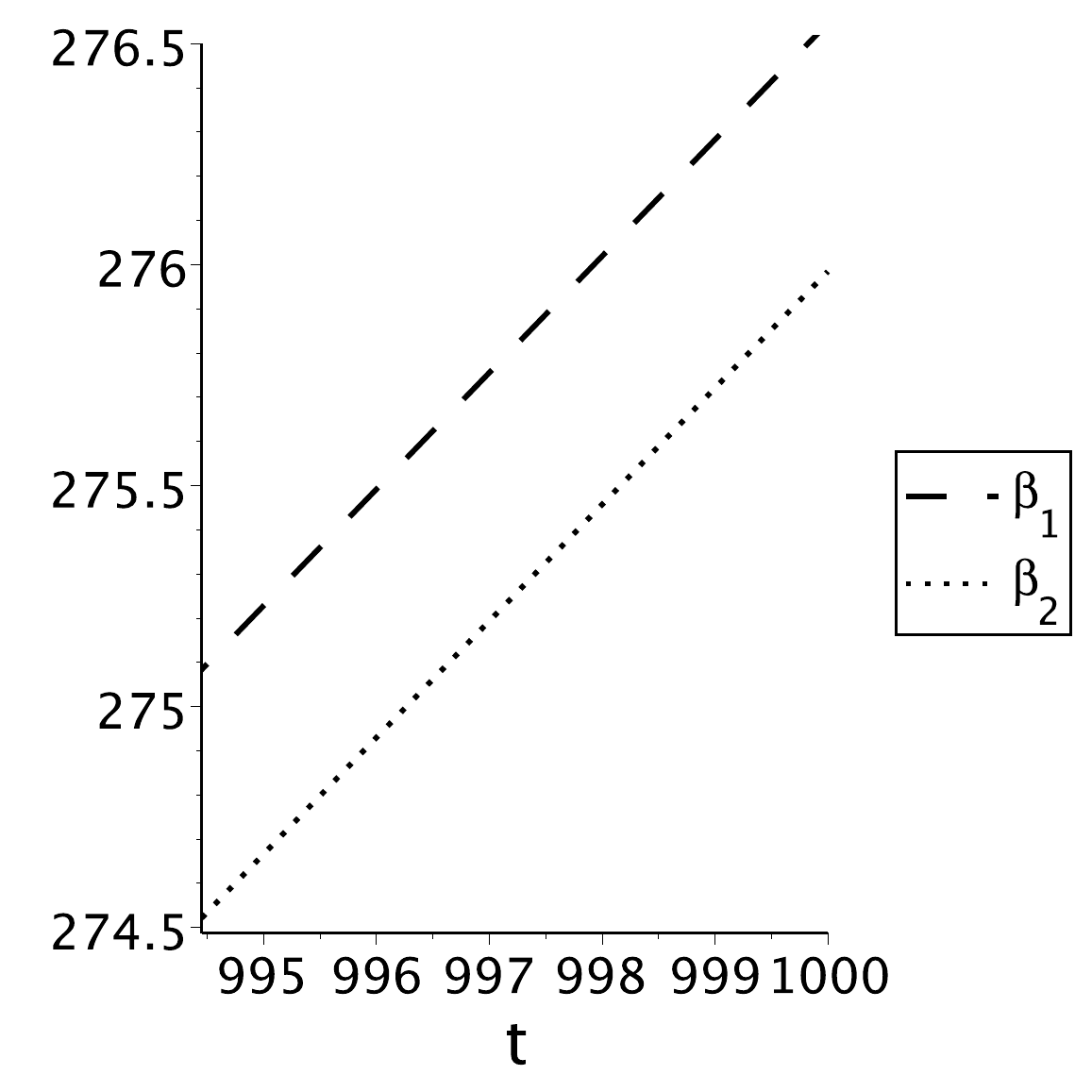}
\caption{Numerical solution to system of equations \eqref{11a}, \eqref{22a} and \eqref{Ra} for initial conditions $ \beta_1(0)=0.5, \ \beta_2(0)=0.1, \ \dot{\beta}_1(0)=0, \ \dot{\beta}_2(0)=0.5, \ \dot{R}(0)=0$. $R(0)\simeq 12.6745229$ is found from equation \eqref{eq00a}. For the found numerical solution $H_1 =H_2 \simeq 0.263569$, $R_0 \simeq 2.917699$.}
\label{fig2}
\end{figure}

Fig.\ref{fig2} shows the numerical solution of the system of equations \eqref{11a}, \eqref{22a} and \eqref{Ra} in various regions of time variation. The beginning of the motion is shown at the left panel. The time behavior of both functions $\beta_1(t)$ and $\beta_2(t)$ differ from each other due to difference in the initial conditions, see the capture of  Fig.\ref{fig2}. The panel in the middle indicates similar asymptotic behavior of the solutions. The more detailed figure on the right panel helps to distinguish the functions $\beta_1(t)$ and $\beta_2(t)$.

\begin{figure}[ht!]
\includegraphics[width=7cm]{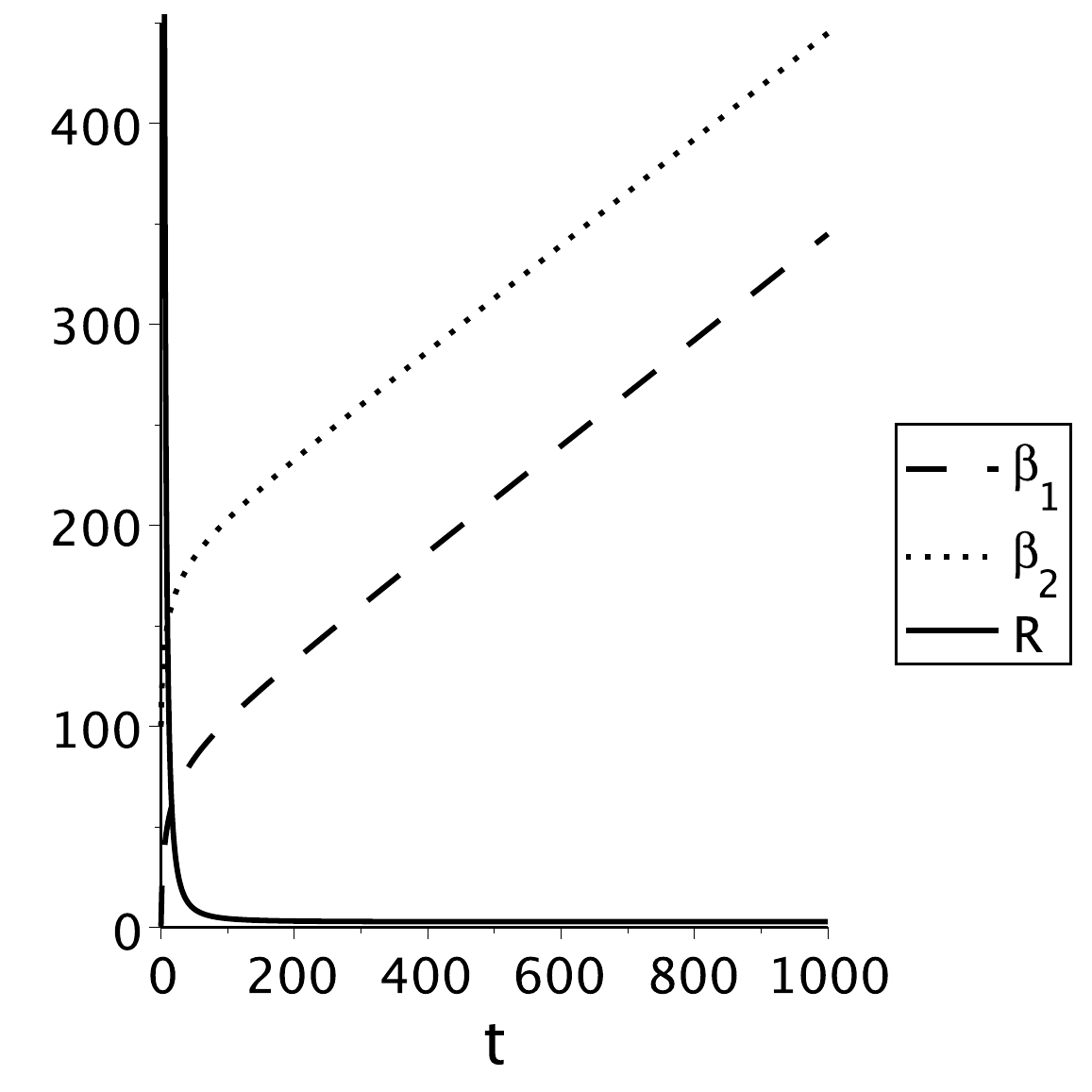}  \quad
\caption{Numerical solution to system of equations \eqref{11a}, \eqref{22a} and \eqref{Ra} for initial conditions $ \ \beta_1(0)=0.5, \ \beta_2(0)=100, \ \dot{\beta}_1(0)=0, \ \dot{\beta}_2(0)=50, \ \dot{R}(0)=0$. $R(0)\simeq 22503.0545$ is found from equation \eqref{eq00a}.}
\label{fig3}
\end{figure}

Variation of the boundary conditions does not alter the asymptotes - see Fig.\ref{fig3}. This observation can be proved analytically.
Suppose that the asymptotic behavior is as follows
\begin{equation} \label{asHH}
	\beta_1(t) = H_1 t,\quad 	\beta_2(t) = H_2 t \quad (H_1>0, \ H_2 >0).
\end{equation}
which is usual for the De Sitter metric. In this case we can strongly simplify the equations of motion. At $t \rightarrow \infty$ equation \eqref{Ra} gives
\begin{equation} \label{Ras1}
R(t)=d_1 (d_1+1)H_1^2 + d_2 (d_2+1)H_2^2 + 2d_1d_2 H_1H_2 \equiv R_0,
\end{equation}
Equations \eqref{11a}, \eqref{22a}, \eqref{00a} at $t \rightarrow \infty$  are transformed into the system of algebraic equations
\begin{eqnarray}\label{space}
\left. f_R \left(d_1 H_1^2 + d_ 2H_1 H_2  \right) -\frac12 f \ \right|_{R=R_0}&=&0,\nonumber \\
\left. f_R \left(d_2 H_2^2 + d_1 H_1 H_2  \right) -\frac12 f \ \right|_{R=R_0}&=&0, \\
\left. f_R \left( d_1 {H_1}^2 +d_2 {H_2}^2 \right) -\frac12 f \ \right|_{R=R_0}&=&0.
\end{eqnarray}
According to these equations, the subspaces are expanded with equal speed,
\begin{eqnarray}\label{ss1}
H_1=H_2= \left. \sqrt{ \frac{f}{2 (d_1 +d_2) f_R }}\ \right|_{R=R_0}.
\end{eqnarray}
If we substitute these expressions in \eqref{Ras1}, then we obtain the equation for $R_0$
\begin{eqnarray}
\left. 2  R f_R - (d_1 +d_2 +1) f \frac{}{}\right|_{R=R_0}=0.
\end{eqnarray}

\begin{figure}[ht!]
\includegraphics[width=7cm]{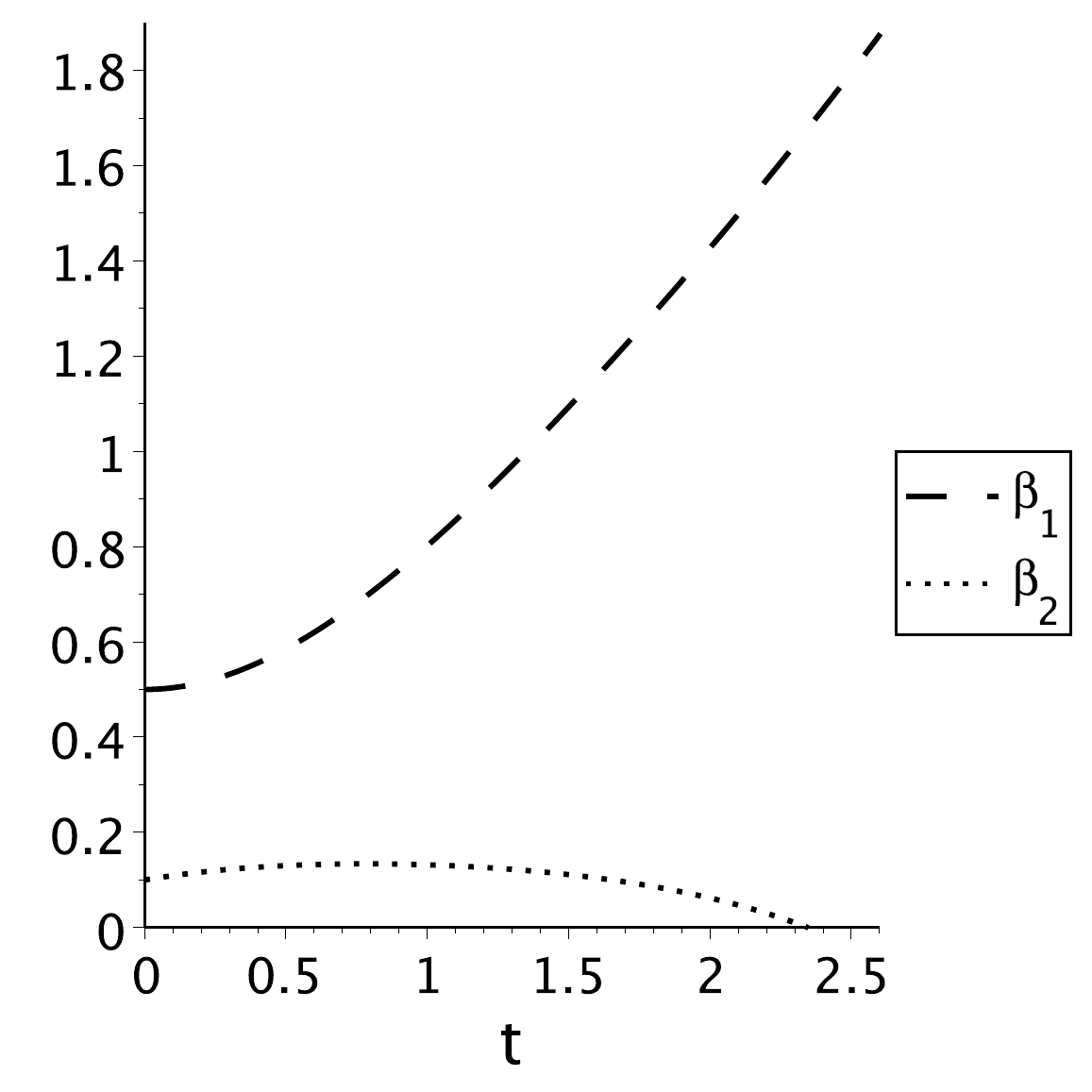}
\caption{The same as in Fig.\ref{fig2} except $\dot{\beta}_2(0)=0.1$}
\label{fig2a}
\end{figure}

We note that the results at $t \rightarrow \infty$  do not depend on the initial conditions and $H_1, H_2, R$ are consistent with the numerical results. Nevertheless, the conclusion that the asymptotic behavior is independent of the initial conditions in their whole range is hasty. Indeed, if we change the parameter $\dot{\beta}_2(0)$ in the set displayed in Fig.\ref{fig2} to the value $\dot{\beta}_2(0)=0.1$ or smaller, the behavior of solution changes drastically. As shown in Fig.\ref{fig2a} the stable solution is absent.

\subsubsection{The case $d_1=3,d_2=2$}

\begin{figure}[ht!]
\includegraphics[width=5cm]{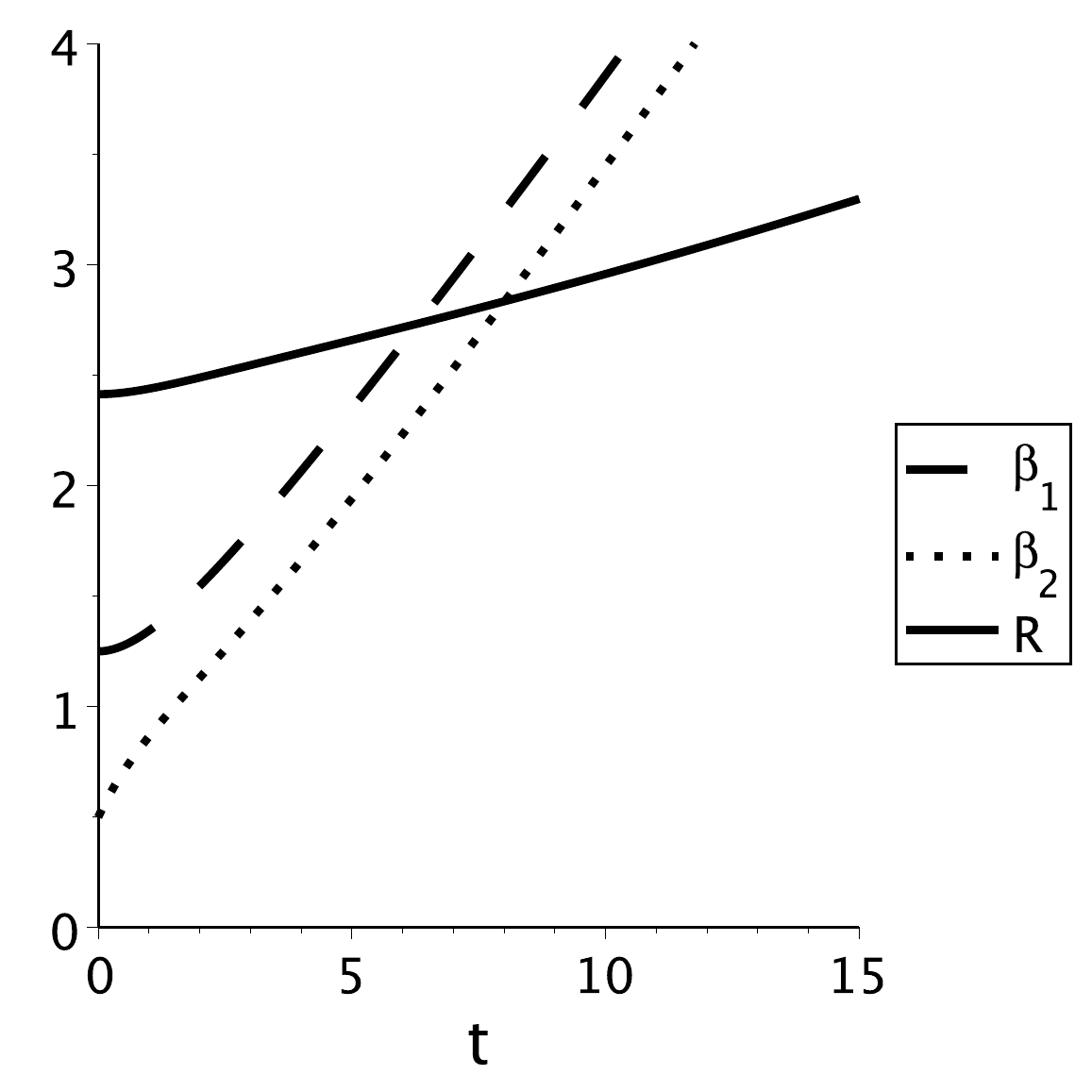}  \quad \includegraphics[width=5cm]{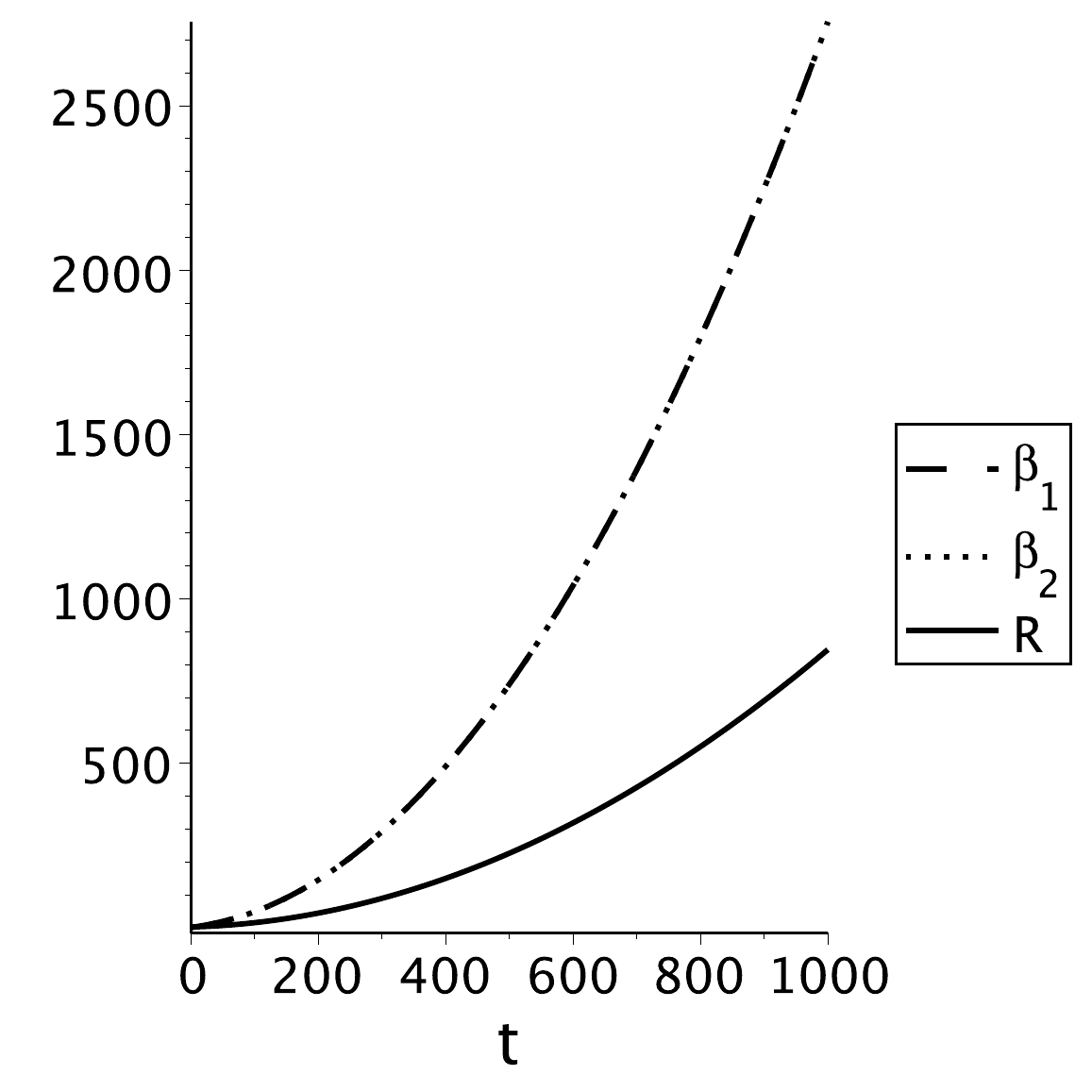}   \quad \includegraphics[width=5cm]{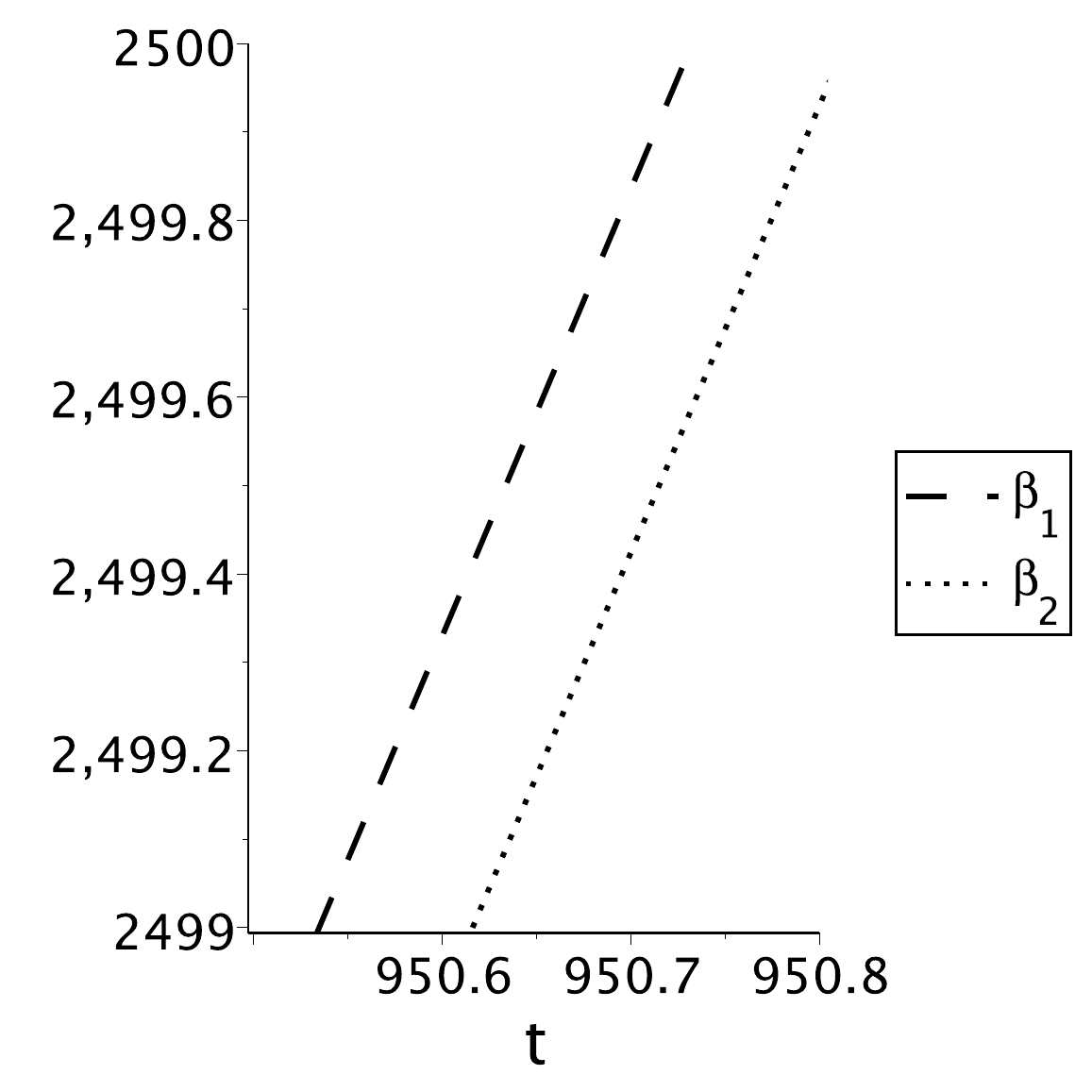}
\caption{The figures show the numerical solution of the system \eqref{11a}, \eqref{22a} and \eqref{Ra} in various regions of time variation for $d_1=3, \ d_2=2$ and initial parameter values $\beta_1(0)=1.25, \ \beta_2(0)=0.5, \ \dot{\beta}_1(0)=0, \ \dot{\beta}_2(0)=0.5, \ \dot{R}(0)=0$. $R(0)\simeq 2.4137960$ is found from equation \eqref{eq00a}.}
\label{fig1}
\end{figure}

Numerical simulation reveals the fact that both subspaces expand equally (see the middle and the right plots in Fig.\ref{fig1}). Moreover, the rate of expansion at $t\rightarrow \infty$ does not depend on the initial conditions.
However, in contrast to the previous case, the expansion rate is proportional to $\exp{(Bt^2)}$.




Let us check analytically that the asymptotes have the form
\begin{equation} \label{as11}
	\beta_1(t) = B_1 t^2,\quad 	\beta_2(t) = B_2 t^2 \quad (B_1>0, \ B_2 >0), \quad t\rightarrow \infty.
\end{equation}
In this case equation \eqref{Ra} gives
\begin{equation}\label{R1}
R(t)=4\left[ d_1 (d_1+1)B_1^2 +2 d_1 d_2 B_1 B_2 + d_2 (d_2+1)B_2^2\right] t^2 +O(t),
\end{equation}
and equations \eqref{11a}, \eqref{22a}, \eqref{00a} look as follows
\begin{eqnarray} \label{12}
&& \hskip-5mm \bigg\{4(d_1 B_1 +d_2 B_2) B_1 f_R  -16\bigg[(d_1 -1) B_1 +d_2 B_2\bigg]\bigg[d_1(d_1+1)B_1^2 +2d_1 d_2 B_1 B_2
\nonumber \\ &&
 \hskip-5mm +d_2(d_2+1)B_2^2 \bigg] f_{RR}
-64 \bigg[d_1(d_1+1)B_1^2 +2d_1 d_2 B_1 B_2 +d_2(d_2+1)B_2^2\bigg]^2 f_{RRR}
\bigg\}t^2
\nonumber \\ &&
 \hskip-5mm +2 B_1 f_R -8\bigg[d_1(d_1+1)B_1^2 +2d_1 d_2 B_1 B_2 +d_2(d_2+1)B_2^2\bigg] f_{RR}
 -\frac{f}{2}=O\left(e^{-2 B_1 t^2}\right),
\end{eqnarray}
\begin{eqnarray} \label{13}
&& \hskip-5mm \bigg\{4(d_1 B_1 +d_2 B_2) B_2 f_R  -16\bigg[d_1 B_1 +(d_2 -1) B_2\bigg]\bigg[d_1(d_1+1)B_1^2 +2d_1 d_2 B_1 B_2
\nonumber \\ &&
 \hskip-5mm +d_2(d_2+1)B_2^2 \bigg] f_{RR}
-64 \bigg[d_1(d_1+1)B_1^2 +2d_1 d_2 B_1 B_2 +d_2(d_2+1)B_2^2\bigg]^2 f_{RRR}
\bigg\}t^2
\nonumber \\ &&
 \hskip-5mm +2 B_2 f_R
-8\bigg[d_1(d_1+1)B_1^2 +2d_1 d_2 B_1 B_2 +d_2(d_2+1)B_2^2\bigg] f_{RR}
 -\frac{f}{2}=O\left( e^{-2 B_2 t^2} \right),
\end{eqnarray}
\begin{eqnarray} \label{14}
&&\bigg\{4(d_1 B_1^2 +d_2 B_2^2) f_R -16\left(d_1 B_1 +d_2 B_2\right)\bigg[d_1(d_1+1)B_1^2 +2d_1 d_2 B_1 B_2
\nonumber \\ &&
+d_2(d_2+1)B_2^2\bigg] f_{RR} \bigg\}t^2  +2(d_1 B_1^2 +d_2 B_2^2) f_R -\frac{f}{2}=0.
\end{eqnarray}
Here we analyze theories with $f(R)$ such that
\begin{equation}\label{ffRR}
\lim_{R \rightarrow \infty}\frac{f}{f_R} \sim R,
\end{equation}
which is true, for example, for theories with a polynomial function
\begin{equation}\label{polinom}
f(R)= a_nR^{n}+a_{n-1}R^{n-1}+... \ ,
\end{equation}
though another class of models that does not satisfy this condition exists, see e.g. \cite{2007PhRvD..75h4010L}.
For our case, expressions
\begin{eqnarray}
&&\lim_{t \rightarrow \infty}\frac{f_R}{f}=\lim_{t \rightarrow \infty}\frac{f_{RR}}{f}
=\lim_{t \rightarrow \infty}\frac{f_{RR}}{f_R}=\lim_{t \rightarrow \infty}\frac{f_{RRR}}{f_R}=0,
\end{eqnarray}
are true if we take into account \eqref{R1}. In turn,
the equations \eqref{12}, \eqref{13}, \eqref{14} acquire the simple form
\begin{eqnarray}\label{space2}
 d_1 B_1^2 +d_2 B_1 B_2 -\lim_{t \rightarrow \infty}\left(\frac{f}{8 t^2 f_R}\right) &=&0, \\
 d_1 B_1 B_2 +d_2 B_2^2 -\lim_{t \rightarrow \infty}\left(\frac{f}{8 t^2 f_R}\right)  &=&0,\\
 d_1 B_1^2 +d_2 B_2^2 -\lim_{t \rightarrow \infty}\left(\frac{f}{8 t^2 f_R}\right)  &=&0.
\end{eqnarray}
It gives
\begin{eqnarray}\label{ss4}
B_1=B_2= \lim_{t \rightarrow \infty}\sqrt{ \frac{f}{8 t^2 (d_1 +d_2) f_R  }},
\end{eqnarray}
and the final expression for the Ricci scalar \eqref{R1} is as follows
\begin{eqnarray}\label{Rpropt2}
R(t)=4 B_1^2 (d_1 +d_2) (d_1 +d_2 +1) t^2 +O(t).
\end{eqnarray}
According to \eqref{polinom} $\frac{\displaystyle f}{\displaystyle f_R} \simeq n R(t) \sim  t^2 $ at $t \rightarrow \infty$. The latter being substituted into \eqref{ss4} gives the relation between the degree of the polynomial "n"\ in \eqref{polinom} and the dimensionality $d_1$ and $d_2$
\begin{equation}
n=\frac12 (d_1+d_2+1)
\end{equation}
The choice made in this subsection $(n=3, d_1=3, d_2=2)$ satisfies this condition. The value of the parameters $B_1 = B_2 = 0.00256$ was obtained numerically.

We conclude that interplay between the dimensionality of extra spaces and the form of the function $f(R)$ influence the asymptotic behavior of the extra space metrics. This conclusion is true for those initial conditions that lead to stable solutions. As was shown numerically, if the derivatives $\dot{\beta}_{1}(0), \dot{\beta}_{2}(0)$ are small there are no stable solutions.

%
%












\subsubsection{Extra space with constant volume, $\beta_2(t)=const$}

The discussion above indicates that time dependence of extra space metric is determined by the initial conditions, though asymptotes have more universal character. The latter leads to growing volumes of both extra spaces. This means that such solutions are hardly applicable to description of our Universe. A more realistic case relates to the situation when one of the subspaces (say $M_{d_2}$) has a stationary radius  $e^{\beta_c}$,
	\begin{equation}\label{bc}
    \beta_2(t)=\beta_c =const,
	\end{equation}
while the other one, $M_{d_1}$, expands.
In this case, the system of equations \eqref{11a}-\eqref{Ra} admits an analytic solution. More definitely, the combination
$d_1 \cdot$ \eqref{11a} $ -d_1 \cdot$\eqref{22a}  +\eqref{00a} $ -f_R \cdot $\eqref{Ra}
gives
	\begin{equation}\label{com}
	-\frac12 f(R)+\left[R(t) +\frac{d_1+d_2-d_2(d_1+d_2)}{e^{\beta_c}} \right]f_R=0.
	\end{equation}
For a given function $f(R)$ this equation determines $R(t)$
	\begin{equation}\label{RR}
	R(t)=R_0=\mbox{const}.
	\end{equation}	
Substituting expressions \eqref{bc} and \eqref{RR} into equations \eqref{11a}-\eqref{Ra} leads to the following system of 4 equations
	\begin{eqnarray} \label{Rc}
	&&R_0 = 2d_1\ddot{\beta}_1 +d_1(d_1+1)\dot{\beta}_1^2 +d_1 (d_1 -1) e^{-2\beta_1(t)} +d_2 (d_2 -1) e^{-2\beta_c} ,
	\end{eqnarray}
    \begin{eqnarray}\label{11c}
	&& -\frac12 f(R_0)+\left. f_R \left[e^{-2\beta_1(t)}(d_1-1)+\ddot{\beta}_1+d_1\dot{\beta}_1^2 \right]\right|_{R=R_0} =0,
	\end{eqnarray}
	\begin{eqnarray}\label{22c}
	&& -\frac12 f(R_0) +\left. e^{-2\beta_c}(d_2-1) f_R\right|_{R=R_0} =0,
	\end{eqnarray}
\begin{equation}\label{00c}
	-\frac12 f(R_0) +\left. d_1\left( \ddot{\beta}_1 +{\dot{\beta}_1}^2 \right) f_R \right|_{R=R_0}=0.
	\end{equation}
Subtracting equation \eqref{11c} from \eqref{00c}, we obtain
	\begin{equation}\label{comc}
	(d_1-1)\left( \ddot{\beta}_1 - \left. e^{-2\beta_1(t)}\right) f_R \right|_{R=R_0}=0,
	\end{equation}
that gives the connection
	\begin{equation} \label{ddb1}
    \ddot{\beta}_1 = e^{-2\beta_1(t)}.
	\end{equation}
for
$ d_1 \neq 1, \ \left. f_R \right|_{R=R_0} \neq 0 $

Then the equations \eqref{Rc} - \eqref{00c} are reduced to
	\begin{equation}\label{eq21}
	\dot{\beta}_1^2 + e^{-2\beta_1(t)} =\frac{(d_2-1) e^{-2\beta_c}}{d_1}=\left. \frac{f(R)}{2 d_1 f_R}\right|_{R=R_0}
    =\frac{R_0 -d_2(d_2-1) e^{-2\beta_c}}{d_1(d_1+1)} \equiv H^2.
	\end{equation}
The solution of equations \eqref{eq21} and \eqref{ddb1} with respect to $\beta_1(t)$ is
	\begin{equation}\label{b1bc}
    {\beta}_1(t) = \mp H\left( t -t_0 \right) +\ln \left( \frac{1 +e^{\pm 2 H (t-t_0)}}{2 H}  \right), \quad H>0,
	\end{equation}
where $H, \ R_0$ and $\beta_c$ can be found from the last relations \eqref{eq21}.
Analytical solution \eqref{b1bc} at $t \rightarrow \infty$ describes the de Sitter space with nonzero Hubble parameter $H$.

The result looks promising, but our numerical simulation indicates that this solution is instable.







In this section we have obtained the set  of asymptotic metrics in the framework of $f(R)$ gravity. Dependence on the initial conditions appears to be nontrivial.
Nevertheless, two issues should be clarified. Namely, the question on the stability of these solutions remains. Connection between the initial conditions and the asymptotes of the metric is also not clear.

\section{Classical evolution of extra metric}

\subsection{Destiny of subspaces and initial conditions}
	
This Section is devoted to the problems mentioned above on more realistic basis. The fact is that $f(R)$ theory with compact maximally symmetric extra dimensions can not reproduce the Minkowski space at low energies \cite{2017JCAP...10..001B}. This means that modern de Sitter stage (which is extremely close to the Minkowski metric) can be achieved by the price of ultra-fine tuning of Lagrangian parameters.  Therefore we have to start with a more general form of action.

The Gauss-Bonnet Lagrangian
\begin{eqnarray}
\label{L_GB}
{\cal L}_{GB} = k\sqrt{-g}\biggr\{R^2 - 4R_{AB}R^{AB} + R_{ABCD}R^{ABCD}\biggl\}
\end{eqnarray}
is the appropriate starting point because of the absence of higher derivatives in the equations of motion. Lagrangian containing the Gauss-Bonnet term plus $f(R)$ term (a function of the Ricci scalar) was used to describe the dark energy phenomenon \cite{2007PhLB..651..224N}. Nevertheless we will use more general form of the pure gravitational action \cite{2006PhRvD..73l4019B}
\begin{eqnarray}\label{Sgen}
&S_{gen}=&\frac{m_D^{D-2}}{2} \int d^D x \sqrt{g_D}
[f(R)+c_1R_{AB}R^{AB}+c_2R_{ABCD}R ^{ABCD}], \\
&& f(R)=aR^2+bR+c.\nonumber
\end{eqnarray}
It is assumed that such Lagrangian is the basis of an effective theory \cite{2007ARNPS..57..329B}. In the following, the parameter $b=1$ without the lost of generality. The Gauss-Bonnet term can be restored if $c_1=-4, c_2 =1$.

In this section we develop the approach to study simultaneous evolution of two subspaces analytically. To make notations more familiar, we will denote $M_{d_0}= T\times M_{d_1},\quad  D=d_0+d_2, d_0=d_1+1$ ($d_0 =4$ for our Universe) and consider metric of the form
\begin{equation}
	ds^2 = g_{AB} dx^A dx^B = g_{ik}(x) dx^{i} dx^k + g_{ab}(x,y) dy^a dy^b,
\end{equation}                                \label{ds1}
where the indexes $i, j, k, \ldots$ refer to the $d_0$-dim part of the metric, and $a,b,\ldots$ to its extra-dimensional part; $(x)$ and $(y)$ mark the dependence on $x^i$ and $y^a$, respectively. The subspaces $M_{d_1}$ and $M_{d_2}$ are assumed to be maximally symmetric.

A substantial simplification of the field equations is achieved if we consider only those events for which the radius $r_1 \equiv e^{2\beta_1(t)}$ of the subspace $M_{d_1}$ is much larger size than the radius $r_2 \equiv e^{2\beta_2(t)}$ of the space $M_{d_2}$, or in terms of the Ricci scalars
\begin{equation}\label{ineq0}
R_{d_1}\ll R_{d_2}.
\end{equation}
We also suppose that the metrics of both spaces $M_{d_1}$ and  $M_{d_2}$ vary slowly along $d_0$ coordinates as compared to the extra coordinates $y^a$. More specifically,
\begin{equation}\label{ep}
   |\partial_{k} g_{AB}| \sim \epsilon |\partial_a  g_{AB}|,
\end{equation}       						
  that is, each derivative $\partial_k$ contains a small parameter $\epsilon$.

This choice breaks the equivalence of these two subspaces and we can follow the method elaborated in \cite{2006PhRvD..73l4019B}. It will be shown that the larger subspace grows constantly while the destiny of the smaller subspace depends on initial conditions.

Then metric \eqref{metric1} leads to the Ricci scalar in the form
	\begin{equation}\label{Tailor}
	R=R_{d_0} + R_{d_2}+ P_k;\quad P_k = 2d_2\ddot{\beta}_2 +d_2(d_2+1)(\dot{\beta}_2)^2 +2d_1d_2\dot{\beta}_1\dot{\beta}_2
	\end{equation}
Additional inequality
	\begin{equation}\label{ineq00}
 P_k\ll R_{d_2}
	\end{equation}
followed from agreement \eqref{ep} means that the functions $\beta_1 (t), \beta_2 (t)$ varies slowly.
According to this inequality and relation \eqref{Tailor} at hand we can perform the Tailor decomposition of the function $f(R)$
	\begin{eqnarray}\label{actTailor}
	&S=&\frac{v_{d_2}}{2}\int d^{d_0}x \sqrt{-g_0}e^{d_2\beta_2} [f'(R_{d_2})R_{d_0} + f'(R_{d_2})P_k + f(R_{d_2})+ \nonumber \\
	&& +c_1R_{AB}R^{AB}+c_2R_{ABCD}R ^{ABCD}+o(\epsilon^4)]
	\end{eqnarray}
in the units $m_D =1$. Here $ v_{d_2}= 2\pi^{\frac{d_2+1}{2}}/\Gamma(\frac{d_2+1}{2})$ is the volume of $d_2$-dimensional sphere.

The Planck mass may be determined as a multiplier of the Ricci scalar $R_{d_0}$ in expression \eqref{actTailor}. It is assumed that the field $\beta_2$ has been settled in a potential minimum in the modern epoch of the Universe evolution and hence $\beta_2 =\beta_c =const$ for the second subspace. In this case the observable Planck mass is
	\begin{equation}\label{MPl}
M_{Pl}^2=v_{d_2}e^{d_2\beta_c}f'(\phi_c).
	\end{equation}
Here the asymptotes $\beta_2(t=\infty)=\beta_c$ and $\phi_c= d_2(d_2-1)e^{-2\beta_c}$ are written in the Jordan frame.
 
It is more familiar to work in the Einstein frame. To this end we have to perform conformal transformation  
\begin{equation}\label{conform}
g_{ik}\rightarrow g_{ik}^{(E)}=e^{d_2\beta_2} |f'(\phi_2)|g_{ik}
\end{equation}
of the metric describing the subspace $M_{d_0}$.
That leads to the action in the Einstein frame in the form \cite{2006PhRvD..73l4019B}
\begin{eqnarray}\label{Sscalar}
&& S_E=\frac{v_{d_2}}{2}  \int d^{d_0} x \sqrt{-g^{(E)}_0}sign(f')[R^{(E)}_{d_0} + K_E(\phi_2)\dot{\phi}^{2}_2 -2V_E (\phi_2) ] \\
&& K_E(\phi_2)=\frac{1}{d_0-2}\left(\frac{-d_2}{2\phi_2}+\frac{f''}{f'}\right)^2 +\left(\frac{f''}{f'}\right)^2+ \frac{d_2}{4\phi^{2}_2}+\frac{c_1+c_2}{f'\phi_2}; \label{KE}\\
&&V_E(\phi_2)=-\frac{sign(f')}{2} \left(\frac{\phi_2}{ d_2(d_2-1)}\right) ^{\frac{d_2}{d_0 - 2}}|f'|^{\frac{-d_0}{d_0-2}} \biggl[f(\phi_2)+\frac{c_V}{d_2}\phi_2^2\biggr];  \label{VE}\\
&&\quad c_V=c_1 + \frac{2c_2}{(d_2-1)}. \nonumber
\end{eqnarray}
We have got Einstein gravity with the uniformly distributed scalar field 
	\begin{equation}\label{phi2}
	\phi_2 (t)\equiv R_{d_2}(t)= d_2(d_2-1) e^{-2\beta_2(t)}
	\end{equation}
where our physical intuition works properly. The extra space volume $v_{d_2}$ plays the role of the Planck mass square in the Einstein frame in $m_D$ units.  Remind for the future that the observable Planck mass \eqref{MPl} is written in terms of the Jordan frame. 

The action \eqref{Sscalar} leads to the Einstein-Hilbert equations with a scalar field. Solutions to these equations, $\beta_1(t)$ and $\phi_2(t)$, evidently depends on initial conditions.  
Due to relation \eqref{phi2}, knowledge of time dependence of $\phi_2(t)$ means knowledge of the metric $\beta_{2}(t)$. Our nearest aim is to study asymptotic behavior of functions $\beta_1(t\rightarrow\infty),\beta_{2}(t\rightarrow\infty)$.

\begin{figure}[ht!]
	\centering
	\includegraphics[scale=0.5]{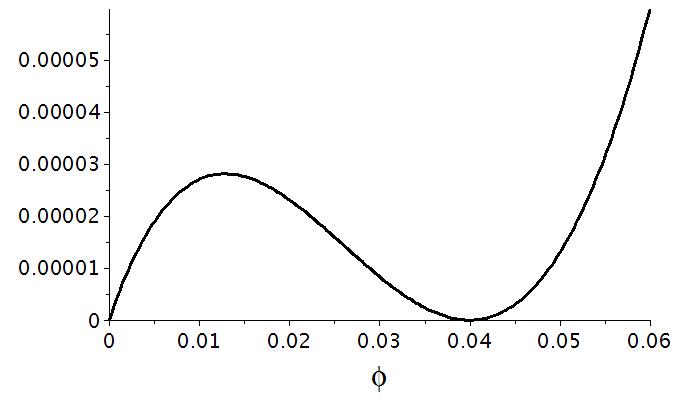}
	\caption{ The potential density in the Einstein frame. The Lagrangian parameters are $d_0=d_1+1=4,d_2 = 2; a = -2; b = 1; c = -0.02; c_1=37, c_2=-32$.}
	\label{VK}
\end{figure}

The potential \eqref{VE} with specific parameter values is represented in Fig. \ref{VK}. We remind that the volume of the subspace $M_{d_1}$ is much greater than the same for $M_{d_2}$ from the beginning, see inequality \eqref{ineq0}.
	
There are two different types of the metrics evolution depending on initial conditions.
The first case is realized when the space $M_{d_2}$ is nucleated with the Ricci scalar $\phi_{2,in}$ from the right of the potential maximum, see Fig.\ref{VK}. The final stationary state is characterized by the field in the potential minimum at $\phi_2=\phi_c=const=0.04$. It means that the Ricci scalar $R_{d_2}\equiv \phi_c =const$ and hence the size of extra space $M_{d_2}$ is also constant, 
\begin{equation}\label{ass1}
r_{d_2}(t)\equiv e^{\beta_{2}(t)},\quad r_{d_2}(t\rightarrow\infty))=e^{\beta_{2}(t\rightarrow\infty)}= \sqrt{\frac{d_2(d_2-1)}{\phi_c}}.
\end{equation} 

On the other hand, the constant term $V_E(\phi_c)$ in the action \eqref{Sscalar} plays the role of the cosmological constant. Therefore, another subspace $V_{d_0}$ is described by the De Sitter metric at $t\rightarrow\infty$ provided that  $V_E(\phi_c)>0$. Its scale factor grows exponentially.
	
The second case is realized when the space $M_{d_2}$ is nucleated with the Ricci scalar $\phi_{2,in}$ from the left of the potential maximum, see Fig.\ref{VK} and the field $\phi_2$ starts its classical motion to zero value. That means that the size of the extra space is increasing with time according to relation \eqref{ass1}.

We conclude that a destiny of nucleated subspaces drastically depends on initial conditions, see also discussion in \cite{2018GrCo...24..154B}. There are only two ways of metric evolution if one subspace is much greater than the other from the beginning.

\subsection{Fixing parameters}
The action \eqref{Sgen} contains several parameters - $c_1, c_2$ and those containing in the function $f(R)$. 
Some connections between them should be imposed if we intend to consider this model as the ordinary scalar field acting in the Minkowski space. The first conditions
	\begin{equation}\label{VMink}
	V_E(\phi_c)=0;\quad V_E'(\phi_c)=0
	\end{equation}
supply the energy density of the Universe be zero. The inequalities
	\begin{equation}\label{ineq}
	V_E''(\phi_c)>0; \quad K_E(\phi_c)>0;
	\end{equation}
are needed for the stability reasons. We also assume that the curvature of extra space is positive and the Planck mass is a real number - see \eqref{MPl}. This leads to the additional inequalities 
\begin{equation}\label{ineq1}
\phi_c >0,\quad f'(\phi_c)>0.
\end{equation}
The algebraic equations \eqref{VMink} together with definition \eqref{VE} give position of the potential minimum
	\begin{equation}\label{cond1}
	\phi_c =-\frac{b}{2a+c_V}
	\end{equation}
and connection between the Lagrangian parameters
	\begin{equation}\label{cond2}
	c=\frac{b^2}{2(2a+c_V)}
	\end{equation}
valid for the Minkowski metric of the space $M_{d_0}$. Here we assume $b\neq 1$ for generality. 

Consider inequalities \eqref{ineq} and \eqref{ineq1} in more detail.
First inequality in \eqref{ineq} gives
	\begin{equation}\label{cond3}
	-sign\left(\frac{c_V b}{2a+c_V}\right) \cdot (2a+c_V)>0 \rightarrow sign (c_V b)<0.
	\end{equation}
	Second inequality in \eqref{ineq} leads to the following expression:
	\begin{equation}\label{cond4}
	24\left(\frac{a}{c_V}\right)^2 + 8\frac{a}{c_V}+3-\frac{c_1+c_2}{c_V}>0.
	\end{equation}
Inequalities
	\begin{equation}\label{cond5}
	\frac{b}{2a+c_V} <0, \quad c_V<0
	\end{equation}
are the result of expressions \eqref{ineq1}.
	
The action parameters must satisfy conditions \eqref{cond1} - \eqref{cond5} for the \eqref{Sgen} model, so that it is realistic. The numerical values of the parameters used in Fig.\ref{VK} do satisfy these conditions.
	
\subsection{Analysis of the metric dynamic}	
As shown above, there are different ways of the metric evolution depending on the initial conditions. In this Section, we analyze the space expansion before the modern horizon has been appeared. Numerical values of the physical parameters are chosen such that they satisfy conditions \eqref{cond1} - \eqref{cond5}. Suppose that the second subspace  $M_{d_2}$ was nucleated with the Ricci scalar $\phi_2 =\phi_{2,in}=0.02$, see Fig.~\ref{VK} while the first extra space  $M_{d_0}$ was nucleated having much larger the Ricci scalar. The dynamic of the field $\phi_2$ is evident due to mechanical analogy: it will move to the potential minimum at $\phi_{2}=\phi_c=0.04$. Our interest is to evaluate the character of this motion as well as the dynamic of the larger subspace $M_{d_0}$.
	
According to \eqref{Sscalar}, the Lagrangian has the form
	\begin{equation}\label{Lagr}
	L=\frac{v_{d_2}}{2} R_{d_0} + \frac{v_{d_2}}{2} K_E(\phi_2)\partial_t\phi_2 \partial^t\phi_2 -v_{d_2}V_E (\phi_2)
	\end{equation}
For our estimations, it is enough to approximate the kinetic term by the constant, $K_E(\phi_{2,in})=K_E(0.02)\simeq 6\cdot 10^{-6}$ and the potential by a linear function $V_E (\phi_2)\simeq V_E (\phi_{2,in})+V'_E (\phi_{2,in})(\phi_2- \phi_{2,in})$. These approximations are valid in a region around $\phi_{2,in}$. In terms of new field 
\begin{equation}\label{chiphi}
\chi=\sqrt{v_{d_2}K_E(\phi_{2,in})}(\phi_2-\phi_{2,in})
\end{equation} the Lagrangian acquires the form
\begin{equation}\label{Lagr2}
	L=\frac{v_{d_2}}{2} R_{d_0} + \frac{1}{2}\partial_t\chi \partial^t \chi -v_{d_2}V_E (\phi_{2,in}) - \sqrt{\frac{v_{d_2}}{K_E(\phi_{2,in})}}V'_E (\phi_{2,in})\chi 
\end{equation}
in the vicinity of $\phi_{2,in}$. Equation of motion for the scalar field is as follows
\begin{equation}\label{eqscal}
\ddot{\chi}(t)+3H\dot{\chi}(t)+\sqrt{\frac{v_{d_2}}{K_E(\phi_{2,in})}}V'_E (\phi_{2,in})=0.
\end{equation}
Following the spirit of the inflationary ideas we use here the slow roll approximation (the second derivative in \eqref{eqscal} is omitted) and the gravitational part is reduced to equation
	\begin{equation}\label{H}
H\simeq \sqrt{\frac{8\pi V_E (\phi_{2,in})}{3}}
\end{equation}
for $d_1=3$ with denotation $H=\dot{\beta}_1(t)$. Therefore, the size of the first extra space grows as 
\begin{equation}\label{size1}
r_1(t)\equiv e^{\beta_1(t)}\simeq e^{Ht}
\end{equation} while the size of the second extra space evolves along the slope of the potential in Fig.~\ref{VK}. Its motion is described by the solution  
\begin{equation}\label{size2}
e^{\beta_2(t)}=\large[\frac{\phi_2(t)}{d_2(d_2-1)}\large]^{-1/2}=\frac{\sqrt{d_2(d_2-1)}}
	{\sqrt{\phi_{2,in}+\frac{\sqrt{v_{d_2}}V'_E}{\sqrt{24\pi V_E K_E}}}t}
\end{equation}
of equation \eqref{eqscal} in the slow role approximation and relations \eqref{phi2}, \eqref{chiphi}. 

The formulas above are written in the units $m_D=1$. Transition to the physical units can be performed by connection \eqref{MPl}  between the observable Planck mass and $D-$dimensional Planck mass $m_D$. The numerical values of the parameters, see capture of Fig.~\ref{VK} gives $m_D\simeq M_{Pl}/26$ where we have used $\phi_2 =\phi_c=0.04$ at the potential minimum and have restored $m_D$.

Now we can estimate the "Hubble" parameter $H$ \eqref{H}. Keeping in mind the value $V_E(\phi_{2,in})\simeq 2.5\cdot 10^{-5}m_D^4$ taken from the potential curve in Fig.~\ref{VK} we obtain $H\sim 3\cdot 10^{-3}M_{Pl}$. The  period of time growing from the scale $m_D^{-1}\sim 10^{-31}$cm to the inflationary scale $10^{-27}$cm is $t\sim 10^4 /M_{Pl}\sim 10^{-39}$sec where the scale factor is assumed to be growing as $\simeq e^{Ht}$. 

A remark is necessary. Up to now we study the rate of expansion in the Einstein picture. In the Jordan picture the expansion rate is different and can be found by application of formula \eqref{conform} to \eqref{size1}
\begin{equation}\label{key}
a_{Jordan}(t)=a(t)\frac{\phi_2(t)^{d_2/2}}{f'(\phi_2(t))}[d_2(d_2 - 1)]^{d_2/2},\quad a(t)\simeq e^{Ht}
\end{equation}
If the field $\phi_2$ moves slowly to its minimum as it is in most inflationary models, exponential grows of the main space remains the same qualitatively while the size of the second extra space tends to constant.

We have considered the first type of the metric evolution. The second type (the subspace $M_{d_2}$ is also increasing) is realized if this subspace is nucleated with a small value of the Ricci scalar (from the left of the potential maximum in Fig.\ref{VK}). To simplify our analysis, consider motion near $\phi_2 =0$. In this case the functions $K_E$ and $V_E$ may be approximated by the first term in the Tailor decomposition,
$$K_E\simeq K_0/\phi^2,\quad V_E\simeq v_0\phi_2^{d_2/(d_0-2)}$$
Substitution of \eqref{phi2} leads to classical equation for $\beta_2(t)$ which can be solved analytically. The result is
\begin{equation}
\beta_2(t\rightarrow \infty)\Rightarrow C_i t
\end{equation}
where $C_i$ is a constant depending on initial conditions. Therefore the size of subspace $M_2$ grows exponentially.

As the result, the first subspace $M_{d_1}$ exponentially grows while the second space $M_{d_2}$ has two variants of evolution. If it was born with the Ricci scalar $R_{d_2}$ from the left of the potential maximum in Fig. \ref{VK}, this subspace is expanded (its radius is proportional to $\sqrt{2/\phi_2}$). If it was born with the Ricci scalar $R_{d_2} =\phi_2$ from the right of the potential maximum, this subspace volume tends to the finite value. 

\section{Funnel as a connection of two subspaces}



Discussion in this Section is devoted to consequences of the most promising case  we have found - one extra space expands while the volume of the second one tends to constant. The particular coordinate could experience expansion or contraction depending on the initial conditions. The question is: what happens if the initial conditions are different in two areas of the manifold $M$? 

Consider a manifold $M$ with topology $T \times M_1 \times M_2 \times M_3$, where $M_1$ is 1-dimensional infinite flat space, and $M_2, M_3$ are 2-dimensional spheres. We study the pure gravitational field action in the form (\ref{Sgen}) with the metric
\begin{equation}
\label{eq:g_fun}
ds^2 =  A(u) dt^2 - A(u)^{-1} du^2 - e^{2\beta_1 (u)} d\Omega_1^2 - e^{2\beta_2 (u)} d \Omega_2^2.
\end{equation}
Here $A(u)$, $\beta_1(u)$ and $\beta_2(u)$ are functions of the Schwarzschild radial coordinate $u$, $-\infty < u < \infty$. The Ricci scalar has the form
$$
R = 2 e^{-2\beta_1}+2e^{-2\beta_2}-A^{\prime \prime} - 4 A^{\prime} \left( \beta_1^{\prime} + \beta_2^{\prime}\right)$$
\begin{equation}
\label{eq:R_sc}
- 2 A \left(3\beta_1^{\prime 2}+3\beta_2^{\prime 2} + 4 \beta_1^{\prime} \beta_2^{\prime} + 2 \beta_1^{\prime \prime} + 2 \beta_2^{\prime \prime} \right),
\end{equation}
with prime denoting differentiation with respect to $u$.
For the Ricci tensor squared we have
\begin{equation}
R_{AB}R^{AB} = \sum_{i = 1}^{6} R_i^{i2},
\end{equation}
where
\begin{eqnarray}
&&R_1^1 = -\frac{A^{\prime \prime}}{2} - A^{\prime} \left( \beta_1^{\prime} + \beta_2^{\prime}\right), \\
&&R_2^2 = -\frac{A^{\prime \prime}}{2} - A^{\prime} \left( \beta_1^{\prime} + \beta_2^{\prime}\right)-2A\left( \beta_1^{\prime 2} + \beta_2^{\prime 2} + \beta_1^{\prime \prime} + \beta_2^{\prime \prime}\right), \nonumber \\
&&R_3^3 = R_4^4 = e^{-2\beta_1} - A \left( 2\beta_1^{\prime 2} + 2 \beta_1^{\prime} \beta_2^{\prime 2} + \beta_1^{\prime \prime} \right) - A^{\prime}\beta_1^{\prime}, \nonumber \\
&&R_5^5 = R_6^6 = e^{-2\beta_2} - A \left( 2\beta_2^{\prime 2} + 2 \beta_1^{\prime} \beta_2^{\prime 2} + \beta_2^{\prime \prime} \right) - A^{\prime}\beta_2^{\prime}; \nonumber
\end{eqnarray}
whereas the Kretschmann scalar is as follows
$$
R_{ABCD}R^{ABCD} = 2 \left[ \left(R^{12}_{12}\right)^2 + \left(R^{12}_{21}\right)^2 + \left(R^{13}_{13}\right)^2 + \left(R^{15}_{15}\right)^2+\right.$$
\begin{equation}
\left. + \left(R^{23}_{23}\right)^2 + \left(R^{25}_{25}\right)^2 + \left(R^{34}_{34}\right)^2 + \left(R^{34}_{43}\right)^2 + \left(R^{56}_{56}\right)^2 + \left(R^{56}_{65}\right)^2 + 2\left(R^{35}_{35}\right)^2 \right],
\end{equation}
where
\begin{eqnarray}
&&R^{12}_{12} = - R^{12}_{21} = -\frac{A^{\prime \prime}}{2}, \qquad R_{13}^{13} = -\frac{A^{\prime}}{2}\beta_1^{\prime}, \qquad R_{15}^{15} = -\frac{A^{\prime}}{2}\beta_2^{\prime}, \\
&&R_{23}^{23} = -\frac{A^{\prime}}{2}\beta_1^{\prime} - A\left( \beta_1^{\prime \prime} + \beta_1^{\prime 2}\right), \qquad
R_{25}^{25} = -\frac{A^{\prime}}{2}\beta_2^{\prime} - A\left( \beta_2^{\prime \prime} + \beta_2^{\prime 2}\right), \nonumber \\
&&R_{34}^{34} = - R_{34}^{34} = e^{-2\beta_1} - A\beta_1^{\prime 2}, \qquad R_{56}^{56} = - R_{56}^{56} = e^{-2\beta_2} - A\beta_2^{\prime 2}, \nonumber \\
&&R_{35}^{35} = -A\beta_1^{\prime}\beta_2^{\prime} \nonumber .
\end{eqnarray}

We are going to describe the structure of the funnel, which is the transition between the domain with (large $ M_2 $/small $ M_1 $) subspaces and the domain with subspaces (large $ M_1 $/small $ M_2 $)  \cite{2016PhLB..759..622R}. It is implied that the Minkowski metric is realized asymptotically which imposes the following restrictions
\begin{equation}\label{A}
A(u \to \pm \infty) \sim 1,
\end{equation}
\begin{eqnarray}\label{b1}
&&\beta_1(u \to + \infty) \sim \ln u, \qquad \beta_1(u \to -\infty) \sim \ln r_0,\\
&&\beta_2(u \to - \infty) \sim \ln u, \qquad \beta_2(u \to +\infty) \sim \ln r_0. \nonumber
\end{eqnarray}
Here $r_0=e^{\beta_c}$ is the radius of extra space at $u\rightarrow \infty$. The value of $r_0 = 1/\sqrt{-c}$ is connected with the physical parameter $c$ according to formulas (\ref{phi2}), (\ref{cond1}), (\ref{cond2}).

Numerical simulations were performed using the Ritz method which means that metric functions $A(u), \beta_1(u), \beta_2(u)$ in (\ref{eq:g_fun}) are approximated by trial functions. Let us choose the trial functions in the form
\begin{equation}
\label{eq:trial}
A(u) = 1 - \frac{\xi_2}{\sqrt{\xi_1^2 + u^2}},
\end{equation}
\begin{displaymath}
e^{\beta_1(u)} = r_0+\frac{1}{2} \left(u+\sqrt[4]{\xi_3^4+u^4}\right),
\end{displaymath}
\begin{displaymath}
e^{\beta_2(u)} = r_0-\frac{1}{2} \left(u-\sqrt[4]{\xi_3^4+u^4}\right),
\end{displaymath}
keeping in mind conditions \eqref{A}, \eqref{b1}. Parameters $\xi_1, \xi_2, \xi_3$ are not independent. Indeed, the integrand in (\ref{Sgen}) has to decrease faster than $1/u$ when $u$ tends to infinity in order to avoid divergences. The determinant grows as $g \sim u^4$
so the content of square brackets in (\ref{Sgen}) should tend to zero faster than $1/u^3$. It can be obtained by setting all the coefficients of terms $u^n,n\ge -3$ equal zero in the Tailor series of the integrand. For terms $u^n,n\ge -2$ it is achieved by the very choice of trial functions (\ref{eq:trial}). For the term $1/u^3$, the coefficient equals zero if the relation
\begin{equation}
\label{eq:cond22}
\left(\frac{2\xi_2}{c}+\xi_1^2\xi_2+3\sqrt{-c}\xi_3^4\right) (1-4c^2)+3c_1(-c)^{3/2} \xi_3^4 = 0
\end{equation}
holds. This relation between $\xi_1, \xi_2$ and $\xi_3$ makes them be dependent. In what follows we consider $\xi_1,\xi_3$ as independent values, and $\xi_2$ as their function. Parameters $\xi_1, \xi_2, \xi_3$ are to be defined by the action minimization. 


Now we proceed the search of approximate metric in the following manner. The action $S(\xi_1,\xi_3)$ appears to be a function of two variables after the substitution of trial functions \eqref{eq:trial} into action \eqref{Sgen}. Classical solutions should satisfy equations
\begin{equation}
\frac{\delta S}{\delta g_{ab}}=0
\end{equation}
or, {according to the Ritz method,} \cite{Review_Ritz}
\begin{equation}\label{appsol}
\frac{\partial S}{\delta \xi_1}=0,\quad \frac{\partial S}{\delta \xi_3}=0.
\end{equation}
One of the ways to solve this system is to find a minimum of the auxiliary function
\begin{equation}
\label{eq:FUN}
\Omega(\xi_1, \xi_2(\xi_1,\xi_3), \xi_3) = \left( \frac{\partial S}{\partial \xi_1} \right)^2 + \left( \frac{\partial S}{\partial \xi_3} \right)^2.
\end{equation}
Note that if the trial functions are chosen absolutely correctly by accident, a minimum of the function $\Omega_{min}=0$. Probability of such accurate choice is almost zero. So we just search for the values of parameters $\xi_1, \xi_2(\xi_1,\xi_3), \xi_3$ which give the minimum of $\Omega$, and consider the resulting trial functions as approximate solutions of \eqref{appsol}.

The described minimization procedure for $\Omega(\xi_1,\xi_2(\xi_1,\xi_3),\xi_3)$ gives the minimum at $\xi_1^{*} = 9.96,\xi_2^*=9.57,\xi_3^*=1.85$ (see Figs \ref{fig:FUN_xi1} and \ref{fig:FUN_xi3}). One can also notice that the minimum is very deep. It means that the trial functions are chosen properly.

\begin{figure}
\centering
\includegraphics[scale=0.4]{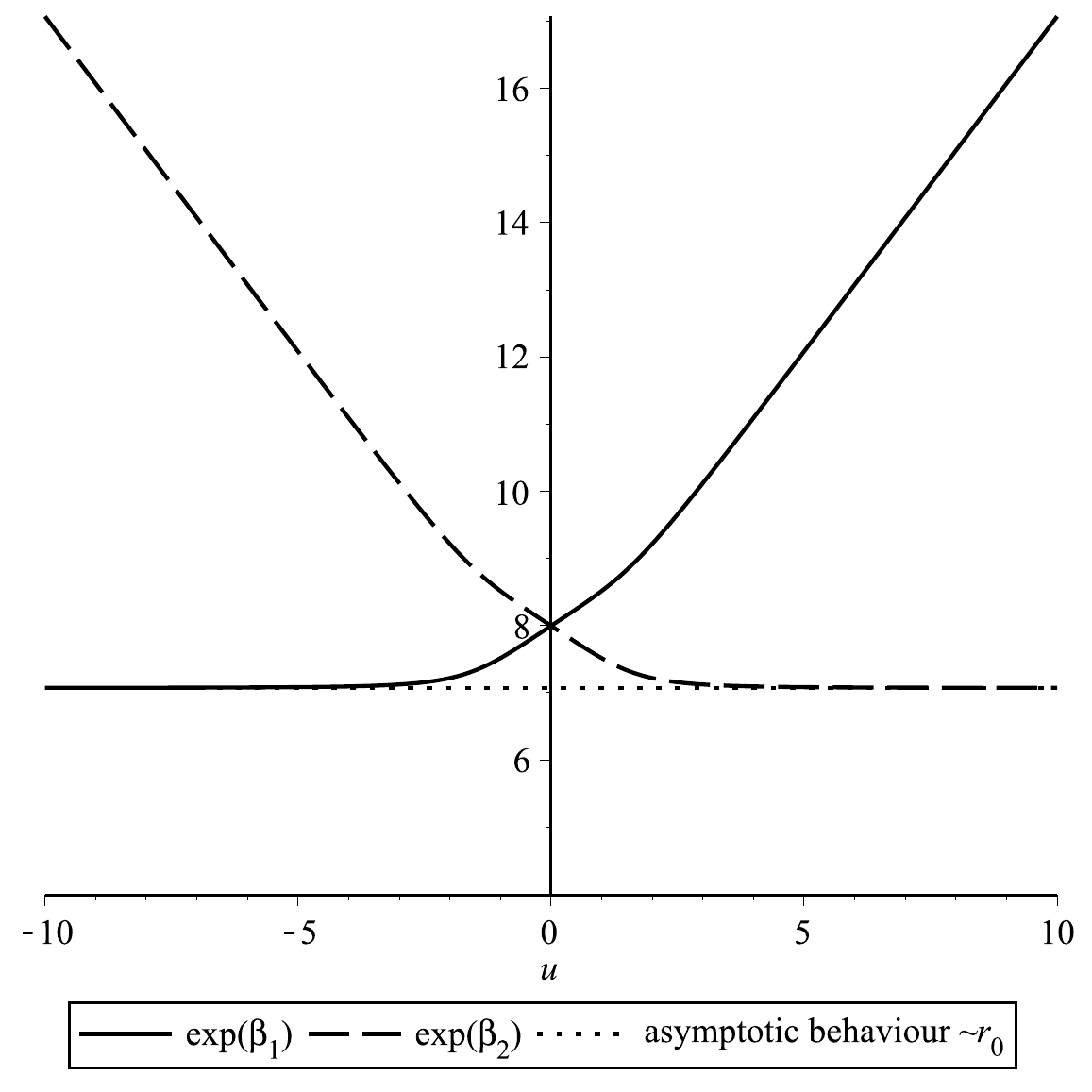}
\caption{Radii of two-dimensional subspaces vs the Schwarzschild radial coordinate $u$ for parameter values  $a = -2, b = 1, c = -0.02$, $\xi_3 = 1.85$.}
\end{figure}

\begin{figure}[ht]
\centering
\includegraphics[scale=0.5]{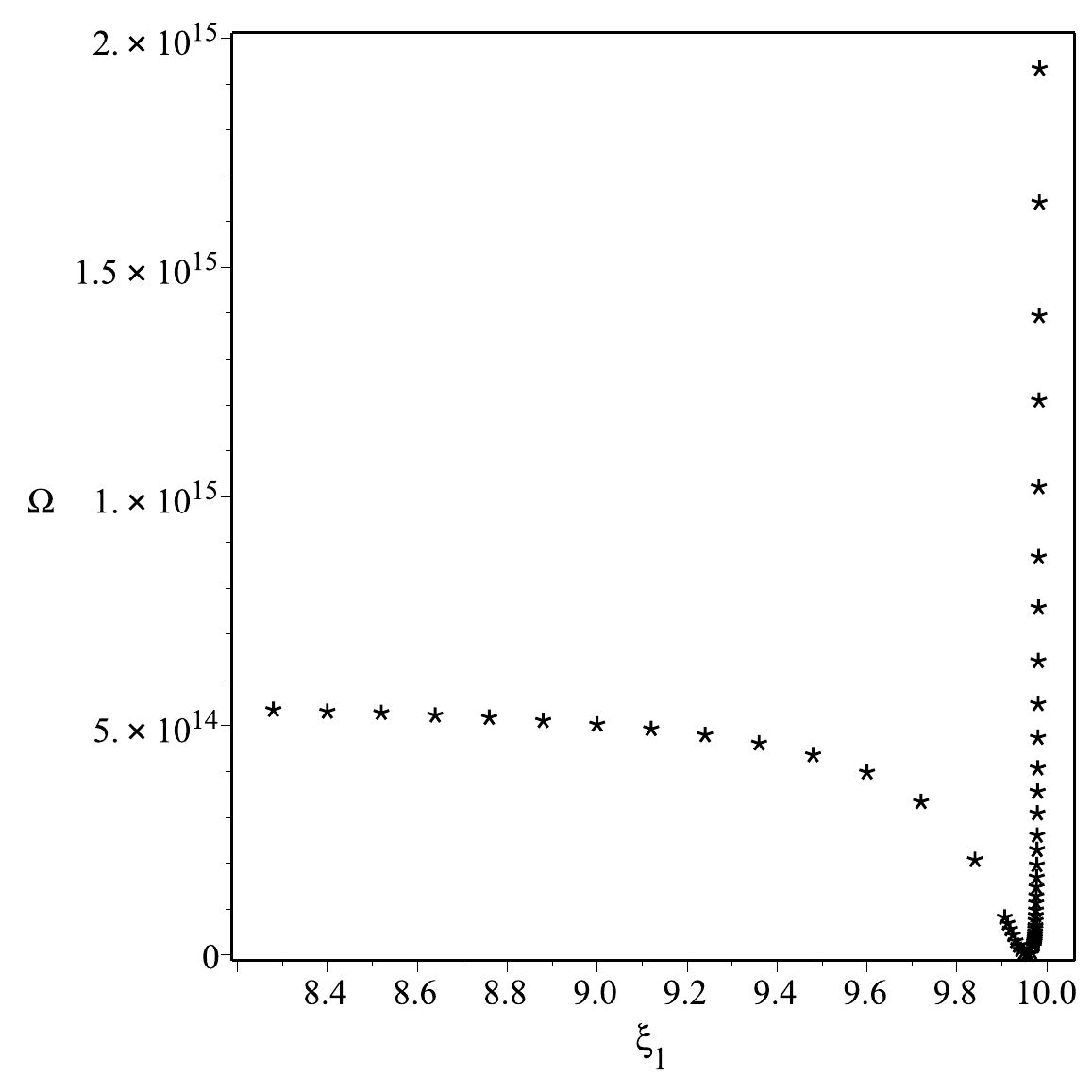}
\caption{Auxiliary function $\Omega$ vs. the minimization parameter $\xi_1$ for $\xi_2^*=9.57,\xi_3^*=1.85$ and $f(R)$-parameters $a = -2, b = 1, c = -0.02$. The minimum of $\Omega$ corresponds to $\xi_1^* = 9.96$. $\Omega (\xi_1^{*} = 9.96, \xi_3^* = 1.85)= 5.3 \times 10^{11}$.}
\label{fig:FUN_xi1}
\end{figure}
\begin{figure}[ht]
\centering
\includegraphics[scale=0.4]{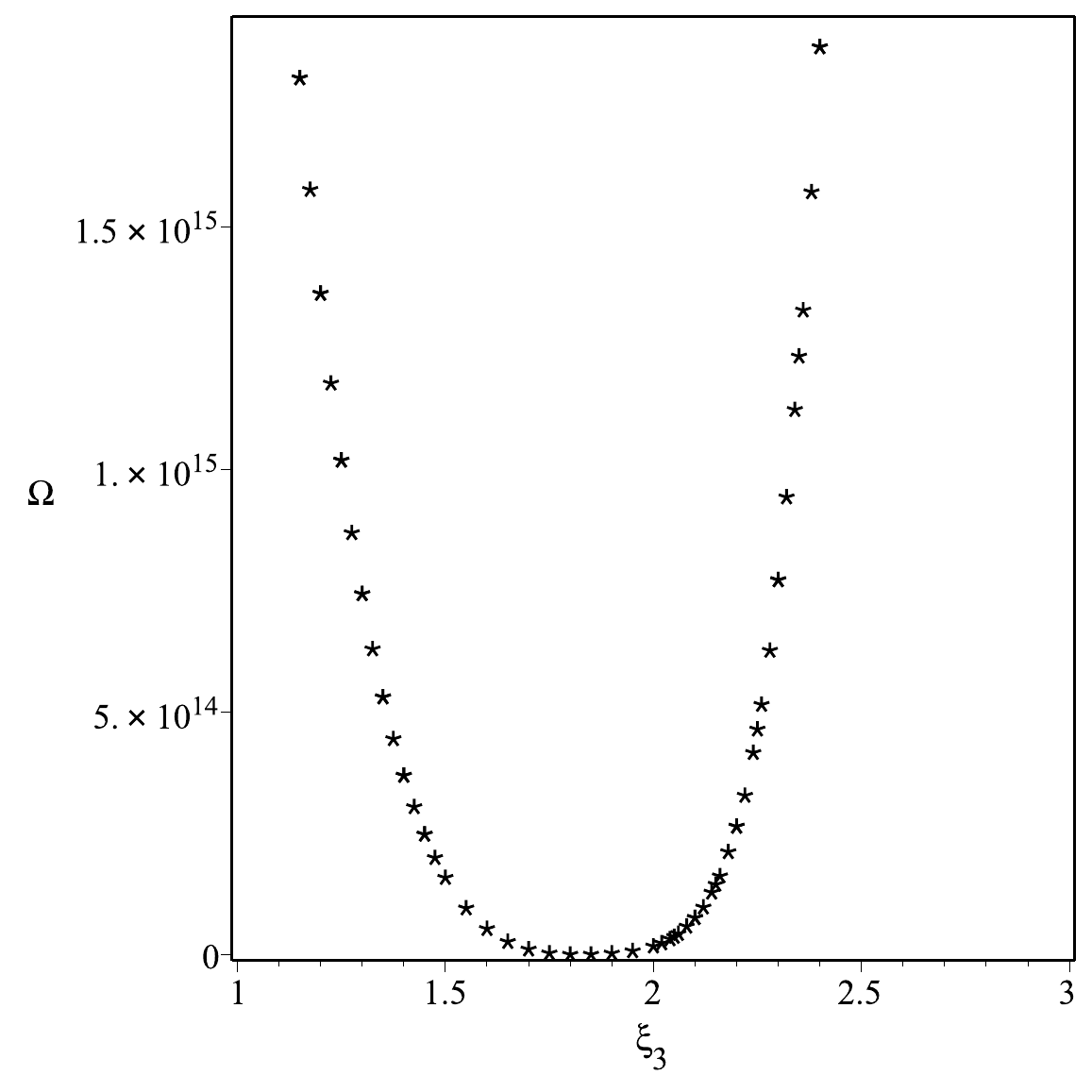}
\caption{Auxiliary function $\Omega$ vs. the minimization parameter $\xi_3$ for $\xi_1^{*} = 9.96,\xi_2^*=1.85$ and $f(R)$-parameters $a = -2, b = 1, c = -0.02$. The minimum of $\Omega$ corresponds to $\xi_3^* = 1.85$. $\Omega (\xi_1^{*} = 9.96, \xi_3^* = 1.85)= 5.3 \times 10^{11}$. }
\label{fig:FUN_xi3}
\end{figure}

\begin{figure}[ht]
\centering
\includegraphics[height=7cm,width=10cm]{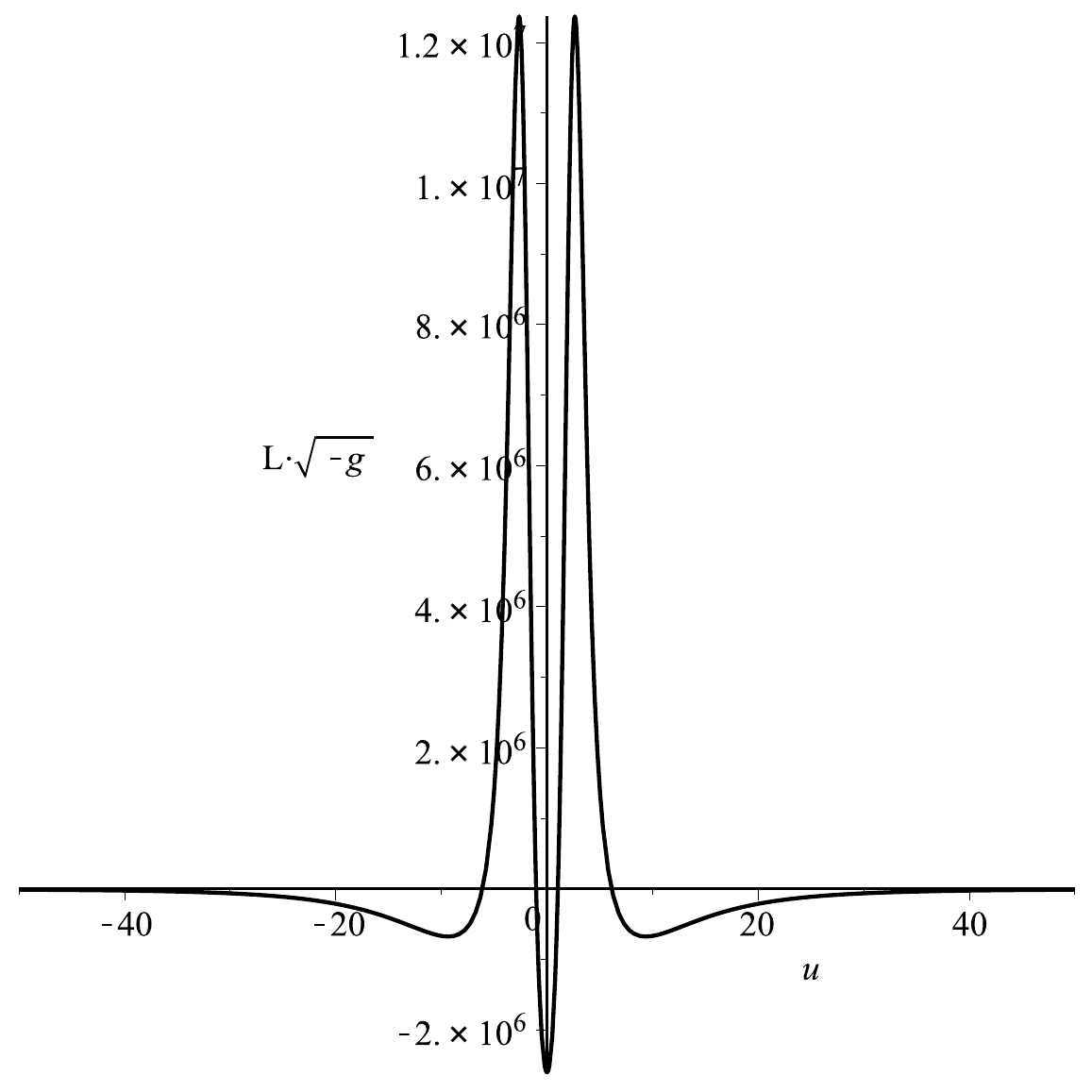}
\caption{Lagrangian density (the integrand in (\ref{Sgen})) vs. the proper distance coordinate $u$. The set of values $\xi_i =\xi_i^*,i=1,2,3$ and $f(R)$-parameters $a = -2, b = 1, c = -0.02$ used.}
\end{figure}

Note that parameters $\xi_2, \xi_3$ have concrete physical sense. The asymptote of $g_{00}$ is $\sim 1 - \xi_2/u$, whereas for weak gravitational field we have $g_{00} \sim 1 - 2m / u$. It means that the distant observer can ''feel'' the parameter $\xi_2$ as the mass of the object (Funnel): $m = \xi_2/2$. The second parameter, $\xi_3$ is related to the size of the funnel transition as $d = r_0 + \xi_3$. Thus it is of interest to present values of these parameters in dimensional form. Using (\ref{MPl}) and the set of values from the capture of Fig. \ref{VK} we obtain
\begin{equation}
m = 0.2 M_{Pl}, \qquad d = 2.7 \times 10^2 M_{Pl}^{-1}.
\end{equation}

For an external observer the funnel looks like a microscopic object with the mass of the order of the Planck scale. It does not interact with matter other than gravitationally and may serve as a dark matter candidate.

\section{Conclusion}

General aim of this research is to analyze the role of initial conditions in asymptotic behavior of subspace metrics. On the basis of pure $f(R)$ gravity we have shown that asymptotic behavior is the same for a wide set of the initial conditions. Exact asymptotes of the subspaces where obtained in an analytical form. On the other hand, the character of asymptotic depends on the topology and dimensionality of space. If the dimensionalities of extra spaces are $d_1 =3, d_2=3$, their sizes are growing as $e^{Ht}$ with the common Hubble parameter $H$ in the Jordan frame. The behavior changes drastically for the case $d_1 =3, d_2=2$. In this case the extra space sizes grow as $e^{Bt^2}$ at $t\rightarrow \infty$. We also obtain analytic result for asymptote of one extra space - formula \eqref{b1} - provided that the size of another one is constant. This result is of purely academic interest. Indeed, as was shown in \cite{2017JCAP...10..001B}, 4-dim Minkowski space-time is incompatible with maximally symmetric compact extra spaces of nonzero curvature in the framework of pure $f(R)$ gravity.

The results based on more complicated model \eqref{Sgen} are more promising. The 4-dim Minkowski space-time and maximally symmetric compact extra spaces with positive curvature could coexist. The price is connection \eqref{cond2} between the parameters of the Lagrangian which may be considered as the strong fine tuning of the model. 
Three kinds of the solutions are found. One of them is characterized by expansion of both subspaces while the second solution describes expansion of one extra space and stabilization of the other. The third solution takes place when the initial conditions are different in separate points of a manifold. In this case, the nontrivial solution (funnel) could be formed. For observers, it looks like a point-like object of the order of the Planck mass and size. They can be considered as the dark matter candidates. 

One can conclude that a number of final states of metric describing our Universe endowed by extra dimensions is quite poor if we limit ourselves with a maximally symmetric extra space. Inclusion of inhomogeneous extra spaces \cite{Rubin:2015pqa,2017JCAP...10..001B,2016arXiv160907361R} improves the situation significantly.

\section{Acknowledgement}

The authors are grateful to Prof. V. Ivashchuk for helpful comments.
The work was supported by the Ministry of Education and Science of the Russian Federation, MEPhI Academic Excellence Project (contract N~02.a03.21.0005, 27.08.2013). 
The work of S.G.R. was also supported by the Ministry of Education and Science of the Russian Federation, Project N~3.4970.2017/BY.
The work of S.G.R. and  A.A.P. is performed according to the Russian Government Program of Competitive Growth of Kazan Federal University.

\section*{Appendix A}

\setcounter{equation}{0}
\renewcommand{\theequation}{A\arabic{equation}}

Here some intermediate formulas used in Section II are presented. There are tree equations for two unknown functions $\beta_1(t), \beta_2(t)$ that have to be analyzed. 	The equations for $(a_1,a_1), (a_2,a_2)$ and (00) components in \eqref{AB} are important. All others are the same or trivial.

	 - $(a_1 a_1)$ component gives
	\begin{equation}\label{11}
	-\frac{1}{2}f(R) + (R_{a_1}^{a_1} +\nabla_{a_1}\nabla^{a_1} - \square) f_R = 0,
	\end{equation}
    
	- $(a_2 a_2)$ component in \eqref{AB} has the form
	\begin{equation}\label{22}
	-\frac{1}{2}f(R) + (R_{a_2}^{a_2} +\nabla_{a_2}\nabla^{a_2} - \square) f_R = 0,
	\end{equation}

	- and $(00)$ component is
	\begin{equation}\label{00}
	-\frac{1}{2}f(R) + (R_{0}^{0} +\nabla_{0}\nabla^{0} - \square) f_R = 0,
	\end{equation}

In the coordinates \eqref{metric1}
	\begin{equation}
	\nabla_{A}\nabla_{B}= \partial_{A}\partial_{B}+\frac12(\partial_t g_{AB})\partial_t \quad ,
	\end{equation}
	\begin{equation}
	\nabla_{A}\nabla^{B}f_R= \nabla_{A}g^{BC}\nabla_{C}f_R=g^{BC}\nabla_{A}\nabla_{C}f_R = g^{BC}[\partial_{A}\partial_{C}+\frac12(\partial_t g_{AC})\partial_t ]f_R.
	\end{equation}
	
 Intermediate formulas for the Ricci tensor
	\begin{eqnarray}\label{Rud}
	&&R_{a_1}^{a_1} =e^{-2\beta_1(t)}\bar{R}_{a_1}^{a_1}+\ddot{\beta}_1+\dot{\beta}_1(d_1\dot{\beta}_1 +d_2\dot{\beta}_2) \\
	&&R_{a_2}^{a_2} =e^{-2\beta_2(t)}\bar{R}_{a_2}^{a_2}+\ddot{\beta}_2+\dot{\beta}_2(d_1\dot{\beta}_1 +d_2\dot{\beta}_2) \\
	&&R_0^0 = d_1\dot{\beta}_1^2 +  d_2\dot{\beta}_2^2+  d_1\ddot{\beta}_1 +  d_2\ddot{\beta}_2
	\end{eqnarray}
	
	\begin{equation}
	\bar{R}_{a_i}^{a_i}=d_i -1
	\end{equation}
	
	\begin{eqnarray}
	&&\nabla_{a_i}\nabla^{a_i}f_R=g^{a_i a_i}[\partial_{a_i}\partial_{a_i}+\frac12(\partial_t g_{a_i a_i})\partial_t ]f_R= \nonumber \\
	&&=g^{a_i a_i}\frac12(\partial_t g_{a_i a_i})\partial_t f_R =\dot{\beta_i}\partial_t f_R ; \quad i=1,2 \\
	&&\nabla_{0}\nabla^{0}f_R=g^{00}[\partial_t\partial_t+\frac12(\partial_t g_{00})\partial_t ]f_R=\partial^2_t f_R
	\end{eqnarray}
	We have kept in mind that the space is the product of two maximally symmetric extra spaces and hence $\partial_{a_1}f_R=0$.
	Choose trace of  \eqref{AB}:
	\begin{equation}\label{trace}
	(d_1+d_2)\Box f_R = -\frac{d_1+d_2+1}{2}f + R f_R
	\end{equation}
	where $\Box = \frac{1}{\sqrt{|g|}} \partial_{A} \sqrt{|g|} g^{AB} \partial_{B}$ is the relativistic D'Alambertian operator,
	\begin{equation}\label{R}
	R=R_A^A = R_0^0 +d_1 R_{a_1}^{a_1} +d_2R_{a_2}^{a_2}
	\end{equation}
	and
	\begin{equation}
	\square f_R=\nabla_{A}\nabla^{A}f_R = \partial^2_t f_R +  (d_1\dot{\beta}_1 +d_2\dot{\beta}_2)\partial_t f_R
	\end{equation}
	
\providecommand{\href}[2]{#2}\begingroup\raggedright\endgroup

\end{document}